\documentclass[12pt, draftclsnofoot, onecolumn]{IEEEtran}

\usepackage[T1]{fontenc}
\usepackage{siunitx}
\usepackage{cite}
\usepackage[cmex10]{amsmath}
\usepackage{amssymb}
\usepackage{makecell}
\usepackage{boldline}
\usepackage{adjustbox}
\usepackage{caption}
\usepackage{tikz}
\usetikzlibrary{shapes,arrows,fit,positioning,shadows,calc}
\usetikzlibrary{plotmarks}
\usetikzlibrary{decorations.pathreplacing}
\usetikzlibrary{patterns}
\usepackage{algorithm}
\usepackage{algpseudocode}
\algnewcommand{\Initialization}[1]{%
  \State \textbf{initialization:}
  \Statex \hspace*{\algorithmicindent}\parbox[t]{.8\linewidth}{\raggedright #1}
}
\usetikzlibrary{automata}
\usepackage{xcolor}

\usepackage{pgfplots}
\pgfplotsset{compat=newest}

\usepackage{graphicx}
\usepackage{ragged2e}
\usepackage{hhline}

\newcommand{\pr}[1]{\ensuremath{\left[#1\right]}}
\newcommand{\pc}[1]{\ensuremath{\left(#1\right)}}
\newcommand{\chav}[1]{\ensuremath{\left\{#1\right\}}}
\newcommand{\PM}[1]{\ensuremath{\left|#1\right|}}

\definecolor{r}{rgb}{0, 0, 0}
\definecolor{r2}{rgb}{0,0,0}

\ifCLASSINFOpdf
\else
\fi
\usepackage{amsmath}
\usepackage[cmintegrals]{newtxmath}
\begin{document}
%

%
\title{Discrete MMSE Precoding for Multiuser MIMO Systems with PSK Modulation}

%
%

\author{Erico~S.~P.~Lopes
				and~{Lukas~T.~N.~Landau,~\IEEEmembership{Member,~IEEE}}
\thanks{E.~S.~P.~Lopes and L.~T.~N.~Landau are with the Centro de Estudos em Telecomunica\c{c}\~{o}es, Pontif\'{i}cia Universidade Cat\'{o}lica do Rio de Janeiro, Rio de Janeiro CEP 22453-900, Brazil, (e-mail: lukas.landau@cetuc.puc-rio.br; erico@cetuc.puc-rio.br).
Parts of this work, have been presented on the 24th International ITG Workshop on Smart Antennas (WSA) in 2020 \cite{MMSE_bb}. 
This work has been supported by the {ELIOT ANR18-CE40-0030 and FAPESP 2018/12579-7} project.
}
}
\maketitle


\begin{abstract}
We propose an optimal MMSE precoding technique using quantized signals with constant envelope. Unlike the existing MMSE design that relies on 1-bit resolution, the proposed approach employs uniform phase quantization and the bounding step in the branch-and-bound method is different in terms of considering the most restrictive relaxation of the nonconvex problem, which is then utilized for a suboptimal design also. 
Moreover, unlike prior studies, we propose three different soft detection methods and an iterative detection and decoding scheme that allow the utilization of channel coding in conjunction with low-resolution precoding. Besides an exact approach for computing the extrinsic information, we propose two approximations with reduced computational complexity.
Numerical simulations show that utilizing the MMSE criterion instead of the established maximum-minimum distance to the decision threshold yields a lower bit-error-rate in many scenarios. Furthermore, when using the MMSE criterion, a smaller number of bound evaluations in the branch-and-bound method is required for low and medium SNR. Finally, results based on an LDPC block code indicate that the receive processing schemes yield a lower bit-error-rate compared to the conventional design.
\end{abstract}



%
\IEEEpeerreviewmaketitle

\section{Introduction}

\IEEEPARstart{M}{ultiuser} Multiple-input multiple-output (MIMO) systems are expected to be vital for the future of wireless communications \cite{6G_Future_Directions}. However, the energy consumption and costs related to using multiple radio frequency front ends (RFFEs) impose a challenge for this kind of  technology \cite{Power_consumption}. \looseness-1
Energy efficiency (EE) is a key requirement for the next generation of wireless communications. According to \cite{6G_Use_cases}, 6G networks will require 10 to 100 times higher EE when compared to 5G, to enable scalable low-cost deployments, with low environmental impact, and better coverage. As stated in \cite{6G_research_challanges,6G_Vision}, another central demand for future networks is higher data reliability.
With this, a challenge for the design of MIMO systems is the reduction of the energy consumption and costs related to the large number of RFFEs with minimum bit-error-rate (BER) compromise.
One approach to realize low energy consumption and low hardware related costs is the consideration of low-resolution data converters. \textcolor{r}{Depending on the pathloss, the energy consumption of data converters can be relevant for the total energy consumption of the RFFE \cite{Amine_2011power}}. Since their consumption scales exponentially with its resolution in amplitude \cite{Walden_1999}, using low-resolution might be favorable. However, the adoption of low-resolution converters can cause significant performance degradation in the BER. 
In this context, it is promising to balance this BER loss using channel coding to approach the requirements of future wireless communications networks \cite{6G_research_challanges,6G_Vision}.
 
\subsection{Related Works}

To mitigate the performance degradation caused by coarse quantization, low-resolution precoding and detection schemes receive increasing attention of the wireless communications community. 
Several precoding strategies with low-resolution data converters exist in literature. {Linear approaches, such as the phase Zero-Forcing (ZF-P) precoder \cite{ZF-Precoding} and the Wiener Filter Quantized (WFQ) precoder \cite{Jacobsson_2017}, benefit from a relatively low computational complexity. However, they yield performance degradation in BER especially for higher-order modulation \cite{M_Joham_ZF,Mezghani2009,one_bit_zf_saxena}}. Therefore, more sophisticated nonlinear approaches  have  been  presented  recently  in \cite{Jacobsson_2017,Squid_precoder,CVX-CIO,Jedda_mmse_mapped,MSM_precoder,L_Chu2019,jacobsson2018nonlinear,Magiq}. However, the methods from \cite{Jacobsson_2017,Squid_precoder,CVX-CIO,Jedda_mmse_mapped,MSM_precoder,L_Chu2019, jacobsson2018nonlinear} imply rounding and the method from \cite{Magiq} implies the convergence to a local minimum, which make the solution provided by these approaches suboptimal in their design criteria. \textcolor{r}{In \cite{Struder_c3po} another suboptimal approach based on a modified MMSE criterion is developed in terms of a practical precoding algorithm for 3-bit phase quantization.}

Moreover, some optimal precoders exist in literature. 
In \cite{Landau2017} a branch-and-bound algorithm was developed for maximizing the minimum distance to the decision threshold (MMDDT) for the case with 1-bit digital-to-analog converters (DACs), which was extended for uniform phase quantization in \cite{General_MMDDT_BB}. 
In addition, in \cite{Jacobsson2018} a branch-and-bound algorithm, is presented for finding the transmit vector that minimizes the mean square error (MMSE) for the case of 1-bit DACs. 

{On the detection side, various different techniques with low-resolution analog-to-digital converters (ADCs) were developed, e.g. \cite{ML_one_bit,mimo_1bit_detection,One_bit_sphere_decoding,Z_shao_2018,Z_Zhang2018,J_choi2016,Wang2015}}. 

\subsection{Main Contributions}
This study proposes both, a precoding and its corresponding detection scheme for a downlink (DL) MIMO system. The base station (BS) is considered to use PSK modulation in conjunction with uniform phase quantizers that have an arbitrary number of quantization regions.

\subsubsection{Precoding scheme}
$\\$
For precoding, we propose a novel branch-and-bound algorithm for finding the vector that yields the minimum MSE. In contrast to the MMDDT criterion considered in \cite{Landau2017,MSM_precoder} and \cite{General_MMDDT_BB}, which is promising in the context of hard decision receivers and the high signal-to-noise ratio (SNR) regime, the used MMSE criterion is more general.

Besides the consideration of phase quantization and PSK modulation, the proposed approach uses a different bounding method as the approach presented in \cite{Jacobsson2018}. 
Whereas the approach in \cite{Jacobsson2018} employs for relaxation the constant envelope constraint, the presented study relies on the convex hull of the discrete non-convex feasible set, which is by definition the most restrictive relaxation for establishing convexity. 

\textcolor{r}{In addition, we propose a suboptimal precoding approach based on the relaxed problem, which we formulate as a convex quadratic program. 
The proposed method is different from the ones presented in \cite{Jacobsson_2017}, \cite{Jedda_mmse_mapped}, \cite{jacobsson2018nonlinear} and \cite{Struder_c3po}. Unlike the approaches from \cite{Jacobsson_2017} and \cite{jacobsson2018nonlinear}, where the relaxation method relies on a modified optimization problem based on the infinity norm, the proposed suboptimal approach, as mentioned before, considers the convex hull of the discrete feasible set. The convex hull formulation in conjunction with the MSE cost function is promising since the MSE objective, in general, tends to solutions with maximum transmit energy which corresponds to the corner points of the polyhedron. Thus, for many cases, the optimal solution of the relaxed problem is already in the original feasible set.} 

\textcolor{r}{Moreover, in the present study, the MSE scaling factor is also part of the optimization problem and, therefore, does not need to be approximated as done in \cite{Jedda_mmse_mapped}. Furthermore, while the method from \cite{Jedda_mmse_mapped} implies QPSK data modulation and $2^q$ quantization phases, the proposed approach allows the entries of the data vector to belong to a PSK modulation and the quantizers to have an arbitrary number of phases.} 

\textcolor{r}{The proposed suboptimal method also diverges from the practical precoding algorithm for 3-bit phase quantization from \cite{Struder_c3po} in terms of the flexibility in the number of quantization phases and the unmodified MSE objective function.
Although it might be possible to learn a suitable parameter adjustment with a neural network for the algorithm in \cite{Struder_c3po} similar to what is proposed in \cite{Studer_NN}, we would like to mention that the proposed approaches do not require parameter adjustment and are compatible with standard solvers.}

{The numerical results show that the proposed branch-and-bound method corresponds to a lower uncoded 
BER in comparison to the state-of-the-art algorithms for the low and intermediate SNR regime.} 
Moreover, the numerical results confirm that when operating in low SNR, only a small number of bounds need to be evaluated to determine the optimal solution. 

\subsubsection{Detection scheme}
$\\$
\textcolor{r}{Although considered as vital for future wireless communications networks, channel coding is hardly discussed in the context of discrete precoding. Soft detection in conjunction with discrete precoding was first considered in \cite{jacobsson2018nonlinear}, where convolutional codes are used and decoded with a BCJR algorithm in the context of OFDM.
}

\textcolor{r}{Different from \cite{jacobsson2018nonlinear}, which for computing the log-likelihood-ratios (LLRs) relies on the common method for AWGN channels, this study proposes three sophisticated approaches that compute extrinsic information considering the effects of the precoder in the probability density function (PDF) of the received signal. The extrinsic information is then used for computing the LLRs via the discrete precoding aware (DPA) iterative detection and decoding (IDD) algorithm.}

While the first method computes the extrinsic information based on the true PDF of the received signal, the second method relies on a nonlinear Gaussian approximation of the original PDF for its computation. Finally, the third relies on a description of the received signal by a linear model with a Gaussian additive distortion term.



{Numerical results show that employing the common LLR computation method for AWGN channels \cite{jacobsson2018nonlinear} causes an error floor in the systems' BER for high-SNR.
By relying on more sophisticated LLR computation methods, the proposed approaches mitigate this problem while also enhancing the overall BER performance of the system.}

\subsection{Remainder and Notation}

The remainder of this paper is organized as follows: Section~\ref{sec:system_model} describes the system model, whereas Section~\ref{sec:Precoding_design} describes a suboptimal algorithm and
an optimal branch-and-bound strategy for the MMSE criterion. Section~\ref{sec:receiver} exposes the receiver design, presents the mentioned soft detectors and the IDD scheme. Section~\ref{sec:numerical_results} presents and discusses numerical results, while Section \ref{sec:conclusion} gives the conclusions.
The convexity analysis and the derivation of the linear model are presented in the appendix.

Regarding the notation, bold lower case and upper case letters indicate vectors and matrices respectively. Non-bold letters express scalars. The operators $(\cdot)^*$, $(\cdot)^T$ and $(\cdot)^H$ denote complex conjugation, transposition and Hermitian transposition, respectively. Real and imaginary part operator are also applied to vectors and matrices, e.g., $\mathrm{Re}\left\{ \boldsymbol{x} \right\} = \left[ \mathrm{Re}\left\{ \left[\boldsymbol{x}\right]_1 \right\},\ldots,    \mathrm{Re}\left\{ \left[\boldsymbol{x}\right]_M \right\}  \right]^T$.

\section{System Model}
\label{sec:system_model}

\begin{figure*} [t]
\centering
\input{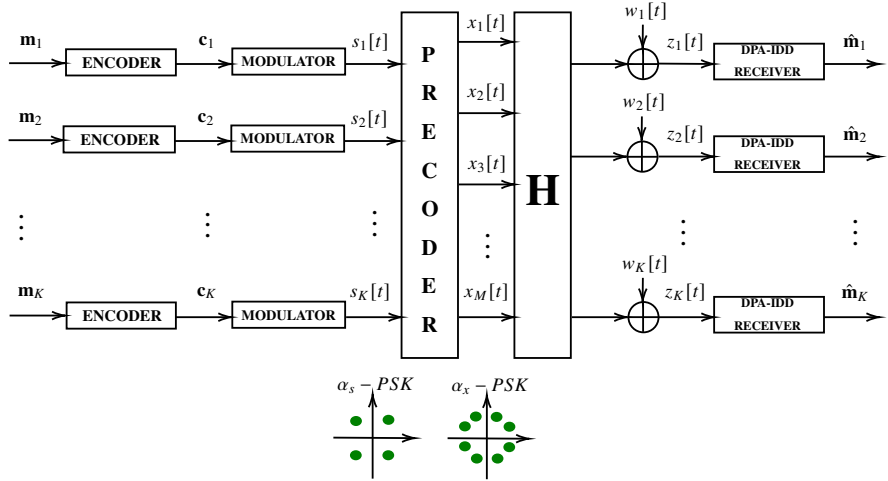}
\caption{Multiuser MIMO downlink with discrete precoding and channel coding}
\label{fig:system_model}       
\vspace{-1em}
\end{figure*}

This study considers a single-cell multiuser MIMO DL system where the BS has perfect channel state information (CSI) and is equipped with $M$ transmitting antennas which serves $K$ single antenna users as  illustrated by Fig.~\ref{fig:system_model}.


A blockwise transmission is considered in which the BS delivers $N_b$ bits for each user. The user specific block is denoted by the vector $\boldsymbol{m}_k=[m_{k,1}, \hdots, m_{k,N_b}]$, where the index $k$ indicates the $k$-th user. 
Each vector $\boldsymbol{m}_k$ is encoded into a codeword vector denoted by $\boldsymbol{c}_k=[{c}_{k,1}, \hdots, {c}_{k,\frac{N_b}{R}}]$, where $R$ is the code rate. {The encoding operation is considered to be systematic meaning that 
$\boldsymbol{c}_k=[{p}_{k,1}, \hdots, {p}_{k,\frac{N_b(1-R)}{R}}, m_{k,1}, \hdots m_{k,N_b}]$, where ${p}_{k,i}$ is the $i$-th parity bit.}

Each encoder provides, sequentially over time slots, $N$ bits to a modulator which maps them into a symbol $s\in \mathcal{S}$ using Gray coding. The set $\mathcal{S}$ represents all possible symbols of a $\alpha_{s}$-PSK modulation, with $\alpha_s=2^N$,  and is described by 
\begin{align}
	    \mathcal{S}= \left\{s: s= e^  \frac{j\pi (2 i+1) }{\alpha_{s}}  \textrm{,  for  }  i=1,\ldots, \alpha_{s} \right\}  \textrm{.}
	    \label{S_set}
\end{align}
The mapping operation is denoted as $s_k[t]=\mathcal{M}\pc{\boldsymbol{r}_{k,t}}$, where $\boldsymbol{r}_{k,t}=[r_{k,t,1}, \hdots, {r}_{k,t,N}]$ is the $t$-th bit vector, taken from $\boldsymbol{c}_k$. The vector $\boldsymbol{r}_{k,t}$ can also be expressed as $\boldsymbol{r}_{k,t}=[c_{k,(t-1)N+1}, \hdots, {c}_{k,t N}]$ for $t=1,\hdots, \frac{N_b}{RN}$.
After mapping, the symbols of all $K$ users are represented in a stacked vector notation as $\boldsymbol{s}[t]=\pr{s_{1}[t] \hdots s_{K}[t]}^{T} \in \mathcal{S}^K $ for each time slot $t$.

The vector $\boldsymbol{s}[t]$ is forwarded to the precoder, which computes the transmit vector $\boldsymbol{x}[t]=[x_{1}[t] \hdots x_{M}[t]]^{T} \in \mathcal{X}^{M}$. The entries of the transmit vector are constrained to the set $\mathcal{X}$, which describes an $\alpha_{x}$-PSK alphabet, denoted by
\begin{align}
	    \mathcal{X}= \left\{x: x= \sqrt{\frac{E_\text{tx}}{M}} \ e^  \frac{j\pi (2 i+1) }{\alpha_{x}}  \textrm{,  for  }  i=1,\ldots, \alpha_{x} \right\} \textrm{,}
	    \label{X_set}
\end{align}
\textcolor{r}{where $E_\text{tx}$ represents the total transmit power. Without loss of generality we consider in this study $E_\text{tx}=1$.}
A frequency flat fading channel is considered, which is described by the matrix $\boldsymbol{H}$ with coefficients $h_{k,m}=g_{k,m}\sqrt{\beta_k}$ where $g_{k,m}$ represents the complex small-scale fading between the $m$-th antenna and the $k$-th user, ${\beta_k}$ denotes the real valued large-scale fading coefficient of the $k$-th user, $k=1...K$ and $m=1...M$.
A block fading model is considered in which the channel is time-invariant during the transmission time.

The BS computes for each coherence time interval of the channel the lookup-table $\mathcal{L}$ containing all possible precoding vectors, which then implies $\boldsymbol{s} \in \mathcal{S}^K \iff \boldsymbol{x}(\boldsymbol{s}) \in \mathcal{L}$.

At the user terminals the received signals are distorted by AWGN denoted by the complex random variable ${w_k}\pr{t}\sim \mathcal{CN}  ({0},\sigma_w^2)$. The receive signal from the $k$-th user is denoted by 
\begin{align}
\label{eq:symbols_at_receiver}
    {z_k}\pr{t} = \boldsymbol{h}_k\ \boldsymbol{x}\pr{t}+{w_k}\pr{t}\text{,}
\end{align}
where $\boldsymbol{h}_k$ is the $k$-th row of the channel matrix $\boldsymbol{H}$. Using stacked vector notation equation \eqref{eq:symbols_at_receiver} can be extended to 
\begin{align}
    \boldsymbol{z}\pr{t} = \boldsymbol{H}\ \boldsymbol{x}\pr{t}+\boldsymbol{w}\pr{t}\text{,}
\end{align}
where $\boldsymbol{z}[t]=\pr{z_{1}[t] \hdots z_{K}[t]}^{T}$, and $\boldsymbol{w}[t]=\pr{w_{1}[t] \hdots w_{K}[t]}^{T}$. 
Each received signal ${z}_k[t]$ is forwarded to the IDD receiver where the transmitted block will be detected. Finally the data block available to the $k$-th user reads as $\hat{\boldsymbol{m}}_k=\pr{\hat{m}_{k,1}, \hdots, \hat{m}_{k,N_b} }$.



\section{MMSE Precoder Design}
\label{sec:Precoding_design}

In this section, we expose the objective of the precoding method and propose two new algorithms for low resolution precoding with phase quantization.
{The proposed precoding approaches do not rely on previous time instants, thus, in this section, we drop the index $t$ notation}.   
\subsection{The Continuous Problem}

With a total transmit energy constraint the MMSE problem can be cast as
\begin{align}
\label{eq:mmse_continuous_problem}
& \min_{\boldsymbol{x},f}  \mathrm{E} \{  \lVert    f{\boldsymbol{z}}  -  \boldsymbol{s}    \rVert_2^2   \} \\
& \text{subject to: }     
  \boldsymbol{x}^H \boldsymbol{x} \leq E_{\textrm{tx}}   \text{,} \ \ \ \ f>0 \text{.}
  \notag
\end{align}
One approach to solve \eqref{eq:mmse_continuous_problem} in closed form is based on KKT conditions and  
the inside that, for the optimal precoding vector, the transmit energy constraint must hold with equality as described in  \cite{M_Joham_ZF}.
Then the optimal precoding vector reads as
\begin{align}
\label{eq:xopt}
 \boldsymbol{x} & =  f^{-1}  \pc{  \boldsymbol{H}^H \boldsymbol{H} +  \frac{ \mathrm{E} \chav{ \boldsymbol{w}^H\boldsymbol{w}  }  } { E_{\text{tx}}}   \boldsymbol{I} }^{-1}                  \boldsymbol{H}^H \boldsymbol{s}  \text{,}
 \end{align}
where the optimal scaling factor is given by
\begin{align}
     f =  \sqrt{    \frac{1}{E_{\text{tx}}}  \pc {   \boldsymbol{s}^H           \boldsymbol{H}
 \pc{  \boldsymbol{H}^H \boldsymbol{H} +  \frac{\mathrm{E} \{ \boldsymbol{w}^H\boldsymbol{w}  \} }   {E_{\text{tx}} }  \boldsymbol{I} }^{-2}                  \boldsymbol{H}^H \boldsymbol{s}  }  
 } \text{.}
 \end{align}
 \textcolor{r}{ The factor $f$ represents a theoretical automatic gain control which is part of the MMSE objective. Since, in this study, PSK modulation is considered, $f$ is not relevant for receive processing.} \looseness-1
\subsection{Problem for Constant Envelope Signals With Phase Quantization at the Transmitter and PSK Modulation}

The problem described in \eqref{eq:mmse_continuous_problem} considers infinite resolution for the entries in $\boldsymbol{x}$. 
Considering quantization of the transmit signal yields the restriction to a discrete input alphabet such that the corresponding problem can be cast as
\begin{align}
\label{LRA_MMSE_precoder}
& \min_{\boldsymbol{x},f}  \mathrm{E} \{  \lVert    f{\boldsymbol{z}}  -  \boldsymbol{s}    \rVert_2^2   \} \\
& \text{subject to: }     
  \boldsymbol{x} \in \mathcal{X}^M\text{,} \ \ \ f>0
  \text{.} 
  \notag
\end{align}


\subsubsection{Proposed Mapped Precoder}
\label{subsec:Proposed_Mapped_Precoder}
$\\$
In this subsection, we propose a suboptimal approach for the problem described in \eqref{LRA_MMSE_precoder}. Since the feasible set of the optimization problem presented by \eqref{LRA_MMSE_precoder} is non-convex we replace
$\mathcal{X}^M$ by its convex hull $\mathcal{P}$, which is a polyhedron. With this, the problem reads as
\begin{align}
\label{eq:relaxed_problem1}
\quad
& \min_{\boldsymbol{x}_{\text{}},f}  \mathrm{E} \{  \lVert  f(\boldsymbol{H}_{\text{}}\boldsymbol{x}_{\text{}} +\boldsymbol{w})  -  \boldsymbol{s}   \rVert_2^2   \} \\
& \text{subject to: }     
 \boldsymbol{x}_{\text{}} \in \mathcal{P} \textrm{,}  \ \ \ f>0 \text{.}  \notag
\end{align}
Rewriting the problem in a real valued notation yields
\begin{align}
\quad
& \min_{\boldsymbol{x}_{\text{r}},f}  \mathrm{E} \{  \lVert  f(\boldsymbol{H}_{\text{r}}\ \boldsymbol{x}_{\text{r}} +\boldsymbol{w}_{\text{r}})  -  \boldsymbol{s}_{\text{r}}\rVert_2^2   \} \\
& \text{subject to: }     
 \boldsymbol{A} \boldsymbol{x}_{\text{r}} \leq \boldsymbol{b} \textrm{,}  \ \ \ f>0  \text{,}  \notag
\end{align}
with
\begin{align}
\label{eq:stacked_vector_notation_for_mmse}
\boldsymbol{x}&_{\text{r}}=		\begin{bmatrix} \mathrm{Re} \left\{\boldsymbol{x}_1\right\} \   \mathrm{Im} \left\{\boldsymbol{x}_1\right\} \
\cdots \
\mathrm{Re} \left\{\boldsymbol{x}_M\right\} \  \mathrm{Im} \left\{\boldsymbol{x}_M\right\}
\end{bmatrix}^T, 
\notag\\
\boldsymbol{w}&_{\text{r}}=   \begin{bmatrix} \mathrm{Re} \left\{\boldsymbol{w}_1\right\} \   \mathrm{Im} \left\{\boldsymbol{w}_1\right\} \
\cdots \
\mathrm{Re} \left\{\boldsymbol{w}_K\right\} \  \mathrm{Im} \left\{\boldsymbol{w}_K\right\}
\end{bmatrix}^T,
\notag\\
\boldsymbol{s}&_{\text{r}}=\begin{bmatrix} \mathrm{Re} \left\{\boldsymbol{s}_1\right\} \   \mathrm{Im} \left\{\boldsymbol{s}_1\right\} \
\cdots \
\mathrm{Re} \left\{\boldsymbol{s}_K\right\} \  \mathrm{Im} \left\{\boldsymbol{s}_K\right\}
\end{bmatrix}^T \notag
\end{align}
and
\begin{align}
\boldsymbol{H}&_{\textrm{r}}    =    \begin{bmatrix} 
\textrm{Re}\chav{{h}_{11}} &-\textrm{Im}\chav{{h}_{11}} 
                         &\cdots&
\textrm{Re}\chav{{h}_{1M}} \ &-\textrm{Im}\chav{{h}_{1M}} 
\\
\textrm{Im}\chav{{h}_{11}} \  &\textrm{Re}\chav{{h}_{11}} 
                         &\cdots&
\textrm{Im}\chav{{h}_{1M}} \  &\textrm{Re}\chav{{h}_{1M}} 
\\
\vdots                         &&\ddots& &\vdots
\\
\textrm{Re}\chav{{h}_{K1}} \ &-\textrm{Im}\chav{{h}_{K1}} 
                         &\cdots&
\textrm{Re}\chav{{h}_{KM}} \ &-\textrm{Im}\chav{{h}_{KM}} 
\\
\textrm{Im}\chav{{h}_{K1}} \ &\textrm{Re}\chav{{h}_{K1}} 
                         &\cdots&
\textrm{Im}\chav{{h}_{KM}} \ &\textrm{Re}\chav{{h}_{KM}} 
\\
\end{bmatrix}  \text{.} \notag
\end{align}

The inequality $ \boldsymbol{A} \boldsymbol{x}_{\text{r}} \leq \boldsymbol{b}$ restricts the elements of the precoding vector to be inside or on the border of the polyhedron whose construction will be detailed in the sequel.
An equivalent problem to \eqref{eq:relaxed_problem1} can be cast as
\begin{align}
& \min_{\boldsymbol{x}_{\text{r}},f}  f^2 \boldsymbol{x}_{\text{r}}^T \boldsymbol{H}_{\text{r}}^T \boldsymbol{H}_{\text{r}}\ \boldsymbol{x}_{\text{r}}  -2 f    \boldsymbol{x}_{\text{r}}^T \boldsymbol{H}_{\text{r}}^T \boldsymbol{s}_{\text{r}}+ f^2 \mathrm{E} \{ \boldsymbol{w}_{\text{r}}^T\boldsymbol{w}_{\text{r}}  \}  \\
& \text{subject to: }     
 \boldsymbol{A} \boldsymbol{x}_{\text{r}} \leq \boldsymbol{b} 
 \text{,}     \ \ \ f>0  \text{.} 
 \notag
\end{align}
If  $f>0$ would be constant, the problem would be a convex quadratic program, since $\boldsymbol{H}_{\text{r}}^T\boldsymbol{H}_{\text{r}} \in S_{+}^{n}$  (cf. Section 4.4 in \cite{Boyd_2004}).
Yet, the problem is in general not jointly convex in $f$ and $\boldsymbol{x}_{\text{r}}$, as can be seen in the Appendix \ref{app:Hessian_MMSE}.

Nevertheless, the problem can be rewritten as an equivalent convex problem by shifting the scaling factor $f$ to the constraints and substituting the optimization variable. This essentially means that the feasible set is scaled depending on the value of $f$.
In this context, we apply the mentioned substitution by introducing a new optimization variable $\boldsymbol{x}_{\text{r,f}} = f\ \boldsymbol{x}_{\text{r}} $, \textcolor{r}{similar as done in \cite{Jacobsson_2017} and \cite{jacobsson2018nonlinear}}.
Accordingly, the resulting problem reads as
\begin{align}
& \min_{\boldsymbol{x}_{\text{r,f}},f}    \boldsymbol{x}_{\text{r,f}}^T  \boldsymbol{H}_{\text{r}}^T \boldsymbol{H}_{\text{r}} \boldsymbol{x}_{\text{r,f}}  -2    \boldsymbol{x}_{\text{r,f}}^T \boldsymbol{H}_{\text{r}}^T \boldsymbol{s}_{\text{r}}+f^2 \mathrm{E} \{ \boldsymbol{w}_{\text{r}}^T\boldsymbol{w}_{\text{r}}  \}  \\
& \text{subject to: }     
 \boldsymbol{A} \boldsymbol{x}_{\text{r,f}} \leq  f \boldsymbol{b} \text{,}  \ \ \ \ \ f>0   \notag  \text{.}
\end{align}
The first constraint can be rewritten as a linear constraint, such that the problem is a convex quadratic program (cf. Section 4.4 in \cite{Boyd_2004}), which reads as
\begin{align}
\label{eq:mapped_mmse_problem}
& \min_{\boldsymbol{x}_{\text{r,f}},f}   \boldsymbol{x}_{\text{r,f}}^T \boldsymbol{H}_{\text{r}}^T \boldsymbol{H}_{\text{r}}\boldsymbol{x}_{\text{r,f}}  -2    \boldsymbol{x}_{\text{r,f}}^T \boldsymbol{H}_{\text{r}}^T \boldsymbol{s}_{\text{r}}+ f^2 \mathrm{E} \{ \boldsymbol{w}_{\text{r}}^T\boldsymbol{w}_{\text{r}}  \}  \\
& \text{subject to: }     
 \boldsymbol{R}
 \begin{bmatrix}
 \boldsymbol{x}_{\text{r,f}}\\
f
\end{bmatrix}
  \leq  \boldsymbol{0},  \ \ \ \ \ f>0  \text{,}  \notag
\end{align}
where $\boldsymbol{R}=  \begin{bmatrix} \boldsymbol{A}   &-\boldsymbol{b} \end{bmatrix}$.
The polyhedron associated to uniformly phase quantized transmit symbols with $\alpha_x$ different phases can be expressed as proposed before in \cite{General_MMDDT_BB}, which is similar to the description in \cite{MSM_precoder}. The corresponding matrix notation reads as
\begin{align}
&\quad\boldsymbol{A}=\begin{bmatrix} (\boldsymbol{I}_M\otimes \boldsymbol{\beta}_1)^T & (\boldsymbol{I}_M\otimes \boldsymbol{\beta}_2)^T& \ldots &(\boldsymbol{I}_M\otimes \boldsymbol{\beta}_{\alpha_x})^T \end{bmatrix}^T  \text{,} \\
&\quad \boldsymbol{\beta}_i=\begin{bmatrix} \cos{\phi_i}& - \sin{\phi_i}\end{bmatrix} ,
\quad  \phi_i=\frac{2\pi i }{\alpha_x}  \textrm{,  for  }  i=1,\ldots \textrm{,}\ \alpha_x  \textrm{,}\\
&\quad\boldsymbol{b}=\frac{\cos(\frac{\pi}{\alpha_x})}{\sqrt{{{M}}}} \boldsymbol{1}_{M\alpha_x} \text{,}
\end{align}
with $\boldsymbol{1}_{M\alpha_x}$ being a column vector with length $M\alpha_x$.
Note that the solution of \eqref{eq:mapped_mmse_problem} yields a lower bound on the optimal value of the original problem, meaning that the corresponding MSE is smaller or equal to the corresponding MSE of the original problem in \eqref{LRA_MMSE_precoder}. Yet, the optimal solution of the relaxed problem is not necessarily in the feasible set of the original problem $\mathcal{X}^M$. 

Therefore, in order to find a feasible solution, mapping to the closest Euclidean distance point in $\mathcal{X}^M$ is considered. The solution after mapping, then, yields a MSE which is always greater or equal to the optimal of \eqref{LRA_MMSE_precoder}, meaning that after the mapping process an upper bound on the optimal value of the original problem is found.

\subsubsection{Proposed Optimal Approach via Branch-and-Bound }
$\\$
As stated before, the continuous solution of \eqref{eq:mapped_mmse_problem} is in general not in $\mathcal{X}^M$ and thus only provides an unfeasible lower bound, or, after mapping,  a feasible upper bound solution for the original problem.
We, then, propose a branch-and-bound strategy that always provides the optimal solution for \eqref{LRA_MMSE_precoder} with significantly reduced computational complexity as compared to exhaustive search. 

%

\begin{figure} 
\centering
\tikzset{every picture/.style={line width=0.75pt}} 

\begin{tikzpicture}[x=0.17pt,y=0.34pt,yscale=-1,xscale=1]

\draw    (736,46) -- (1288,138) ;
\draw    (736,46) -- (184,138) ;
\draw    (736,46) -- (920,138) ;
\draw    (736,46) -- (552,138) ;
\draw    (184,138) -- (322,230) ;
\draw    (184,138) -- (230,230) ;
\draw    (184,138) -- (138,230) ;
\draw    (184,138) -- (46,230) ;
\draw    (552,138) -- (690,230) ;
\draw    (552,138) -- (598,230) ;
\draw    (552,138) -- (506,230) ;
\draw    (552,138) -- (414,230) ;
\draw    (920,138) -- (1058,230) ;
\draw    (920,138) -- (966,230) ;
\draw    (920,138) -- (874,230) ;
\draw    (920,138) -- (782,230) ;
\draw    (1288,138) -- (1426,230) ;
\draw    (1288,138) -- (1334,230) ;
\draw    (1288,138) -- (1242,230) ;
\draw    (1288,138) -- (1150,230) ;
\draw  [dash pattern={on 4.5pt off 4.5pt}]  (-50,138) -- (309.5,138) -- (1429,138) ;

\draw (150,125) node  [scale=0.7] [align=left]    {$x_{1}$};
\draw (530,125) node  [scale=0.7] [align=left]    {$x_{2}$};
\draw (940,125) node  [scale=0.7] [align=left]    {$x_{3}$};
\draw (1300,125) node  [scale=0.7] [align=left]    {$x_{4}$};
\draw (10,225) node  [scale=0.7] [align=left]    {$x_{1}$};
\draw (105,225) node  [scale=0.7] [align=left]    {$x_{2}$};
\draw (190,225) node  [scale=0.7] [align=left]    {$x_{3}$};
\draw (275,225) node  [scale=0.7] [align=left]    {$x_{4}$};
\draw (380,225) node  [scale=0.7] [align=left]    {$x_{1}$};
\draw (475,225) node  [scale=0.7] [align=left]    {$x_{2}$};
\draw (560,225) node  [scale=0.7] [align=left]    {$x_{3}$};
\draw (640,225) node  [scale=0.7] [align=left]    {$x_{4}$};
\draw (750,225) node  [scale=0.7] [align=left]    {$x_{1}$};
\draw (845,225) node  [scale=0.7] [align=left]    {$x_{2}$};
\draw (930,225) node  [scale=0.7] [align=left]    {$x_{3}$};
\draw (1010,225) node  [scale=0.7] [align=left]    {$x_{4}$};
\draw (1115,225) node  [scale=0.7] [align=left]    {$x_{1}$};
\draw (1210,225) node  [scale=0.7] [align=left]    {$x_{2}$};
\draw (1295,225) node  [scale=0.7] [align=left]    {$x_{3}$};
\draw (1380,225) node  [scale=0.7] [align=left]    {$x_{4}$};
\draw (16,74.4) node  [scale=0.7] [align=left]    {$\textcolor{r}{d=1}$};
\draw (16,171.4) node  [scale=0.7] [align=left]    {$\textcolor{r}{d=2}$};

\end{tikzpicture}
\caption{Tree representation of the set $\mathcal{X}^M$ for a system with $M=2$ BS antennas and QPSK precoding modulation ($\alpha_x=4$)}
\label{fig:tree}       
\end{figure}
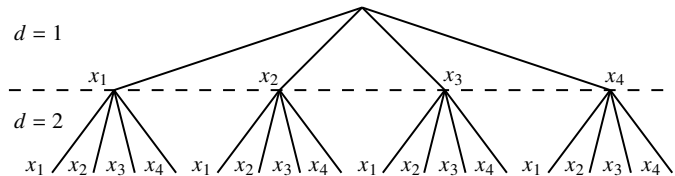

\paragraph{Introduction of the Branch-and-Bound Method}
\label{par:Introduction of the Branch-and-Bound method}
$\\$
A branch-and-bound algorithm is a tree search based method. The tree represents the set of all possible solutions for the vector $\boldsymbol{x}$, i.e., it is a representation of the set $\mathcal{X}^M$. For the construction of the tree $M$ levels are considered and each node has one ingoing branch and $\alpha_x$ outgoing branches as shown in Fig.~\ref{fig:tree}. 

For constructing the precoding vector we consider the minimization of an objective function $g(\boldsymbol{x}, \boldsymbol{s})$, which could be the MSE, subject to the feasible discrete set, described by
\begin{align}
\label{eq:original_problem}
\boldsymbol{x}_{\textrm{opt}} =& \arg\min_{\boldsymbol{x}} g(\boldsymbol{x}, \boldsymbol{s} ) \ \ \textrm{    s.t. } \boldsymbol{x} \in   \mathcal{X}^{M} \textrm{.}
\end{align}

A lower bound on $g(\boldsymbol{x}_{\textrm{opt}},\boldsymbol{s})$ can be obtained by relaxing $\mathcal{X}^M$ to its convex hull. 
The relaxed problem is expressed as
\begin{align}
\label{eq:relaxed_problem}
\boldsymbol{x}_{\textrm{lb}} =& \arg\min_{\boldsymbol{x}} g(\boldsymbol{x}, \boldsymbol{s} ) \ \ \textrm{    s.t. } \boldsymbol{x} \in   \mathcal{P} \textrm{.}
\end{align}
An associated upper bound on $g(\boldsymbol{x}_{\textrm{opt}},\boldsymbol{s})$ can be obtained by mapping the solution of \eqref{eq:relaxed_problem} to the feasible set and evaluating $g(\cdot)$, as discussed previously on subsection \ref{subsec:Proposed_Mapped_Precoder}. The upper bound value of \eqref{eq:original_problem} is termed $\check{g}$.  
Having an upper bound solution implies that $\check{ g } \geq  g(\boldsymbol{x}_{\textrm{opt}}) \geq  g(\boldsymbol{x}_{\textrm{lb}})$, which means that the mapped solution is always greater or equal to the relaxed one from \eqref{eq:relaxed_problem}. 

By fixing $d$ entries of $\boldsymbol{x}$, the vector can be rewritten as $\boldsymbol{x}=[\boldsymbol{x}_1^T, \boldsymbol{x}_2^T ]^T$, with $\boldsymbol{x}_1 \in \mathcal{X}^d $. With this, a subproblem can be formulated as
\begin{align}
\label{eq:lb_subproblem}
\boldsymbol{x}_{2,\text{opt}|\boldsymbol{x}_1} =& \arg\min_{\boldsymbol{x}_2} g(\boldsymbol{x}_2, \boldsymbol{x}_1, \boldsymbol{s} ) \\
&\textrm{    s.t. } \boldsymbol{x}_2 \in   \mathcal{X}^{M-d} \textrm{.} \notag
\end{align}
Relaxing the problem from \eqref{eq:lb_subproblem} we have
\begin{align}
\label{eq:lb_subproblem_relaxed}
\boldsymbol{x}_{2,\textrm{lb}}=& \arg\min_{\boldsymbol{x}_2} g(\boldsymbol{x}_2, \boldsymbol{x}_1, \boldsymbol{s} ) \\
&\textrm{    s.t. } \boldsymbol{x}_2 \in   \mathcal{J} \textrm{,} \notag
\end{align}
where $\mathcal{J}$ is the convex hull of $\mathcal{X}^{M-d}$.

If the optimal value of \eqref{eq:lb_subproblem_relaxed} is larger than a known upper bound $\check{ g }$ on the solution of \eqref{eq:original_problem}, then all members in the discrete set which include the previously fixed vector $\boldsymbol{x}_{1}$ can be excluded from the search process.

By this strategy we intend to exclude most of the candidates from the possible solution set, such that the number of residual candidates is only a small fraction of its total number and, thus, they can be evaluated via exhaustive search.
\paragraph{Branch-and-Bound Initialization}
$\\$
The branch-and-bound algorithm converges faster when we can compute as early as possible an upper bound that permits many exclusions. Therefore, it is recommended to have an initialization step where an upper bound $\Check{g}<\infty$ is found before beginning with the search process. 

In this regard, for initialization, the problem described in \eqref{eq:mapped_mmse_problem} is solved. With this, $\boldsymbol{x}_{\textrm{lb}}$ and $g(\boldsymbol{x}_{\textrm{lb}})=\textrm{MSE}_\textrm{lb}$ are obtained. After mapping $\boldsymbol{x}_{\textrm{lb}}$ to the feasible set $\boldsymbol{x}_{\textrm{ub}}$ and $\Check{g}=\textrm{MSE}_\textrm{ub}$ are determined. 
Note that if the continuous solution of \eqref{eq:mapped_mmse_problem} is in the feasible set, upper and lower bound are equal which can be expressed as
\begin{align}
\label{eq:stopping_condition}
    \boldsymbol{x}_{\textrm{ub}}=\boldsymbol{x}_{\textrm{lb}}=\boldsymbol{x}_{\textrm{opt}} \xrightarrow{}
    g(\boldsymbol{x}_{\textrm{lb}})=\Check{g}  
 \textrm{.}
\end{align}
This would mean that the optimal solution is found already by the approach from subsection \ref{subsec:Proposed_Mapped_Precoder} and the tree search process can be skipped.

\paragraph{Subproblems}
$\\$
When the  condition from \eqref{eq:stopping_condition} is not met, one can apply the branch-and-bound tree search method. It is, then, necessary to solve subproblems, as first mentioned on \ref{par:Introduction of the Branch-and-Bound method}. The equations that define the subproblems are derived below.
First the precoding vector is divided in a fixed vector of length $2d$ and a variable vector according to
 \begin{align}
\boldsymbol{x}_{\text{r}}= \pr{ \boldsymbol{x}_{\text{r,fixed}}^T \ \textrm{,} \ {\boldsymbol{x}^\prime _{\text{r}}}^{T} }^T \text{.}
 \end{align}
 With this, the MMSE problem formulation reads as
 \begin{align}
& \min_{{\boldsymbol{x}^\prime _{\text{r}}},f^{\prime}}  \mathrm{E} \{  \lVert  f^{\prime}(
\boldsymbol{H}_{\text{r}} \pr{
\boldsymbol{x}_{\text{r,fixed}}^T \ \textrm{,} \ {\boldsymbol{x}_{\text{r}}^\prime}^{T}
}^T 
+\boldsymbol{w}_{\text{r}})  -  \boldsymbol{s}_{\text{r}}    \rVert_2^2   \} \notag\\
& \text{subject to: }     
 \boldsymbol{A}^{\prime} {\boldsymbol{x}^\prime _{\text{r}}} \leq \boldsymbol{b}^{\prime}\text{,} \ \ \ f^{\prime}>0  \text{,}     
\end{align}
where $ \boldsymbol{A^\prime} {\boldsymbol{x}^\prime _{\text{r}}} \leq \boldsymbol{b^\prime}$ restricts the elements of the precoding vector to be inside of the set $\mathcal{J}$ and will be detailed in what follows.
The channel can be rewritten accordingly as $\boldsymbol{H}_{\text{r}}=\pr{ \boldsymbol{H}_{\text{r, fixed}} \ \ \boldsymbol{H}_{\text{r}}^{\prime}  }$. Then the problem can be cast as
\begin{align}
& \min_{{\boldsymbol{x}^\prime _{\text{r}}},f^{\prime}}  \mathrm{E} \{  \lVert  f^{\prime} (
\boldsymbol{H}_{\text{r}}^{\prime}\  {\boldsymbol{x}^\prime _{\text{r}}}
+
\boldsymbol{H}_{\text{r, fixed}}\
\boldsymbol{x}_{\text{r, fixed}}
 +\boldsymbol{w}_{\text{r}})  -  \boldsymbol{s}_{\text{r}}    \rVert_2^2   \} \notag \\
& \text{subject to: }     
\boldsymbol{A}^{\prime} {\boldsymbol{x}^\prime _{\text{r}}}\leq \boldsymbol{b}^{\prime} \text{,}  \ \ \ f^{\prime}>0  \text{,}  
\end{align}
and an equivalent problem is given by
\begin{align}
\label{eq:partial_problem}
& \min_{{\boldsymbol{x}^\prime _{\text{r}}},f^{\prime}}  
 \lVert  f^{\prime} 
\boldsymbol{H}_{\text{r}}^{\prime} \ {\boldsymbol{x}^\prime _{\text{r}}}
  -  \boldsymbol{s}_{\text{r}} +  f^{\prime}  \boldsymbol{H}_{\text{r, fixed}}\ 
\boldsymbol{x}_{\text{r, fixed}}        \rVert_2^2 + f^{\prime}{}^2 
   \mathrm{E} \{  \boldsymbol{w}_{\text{r}}^T
   \boldsymbol{w}_{\text{r}}
  \} 
 \\
& \text{subject to: }     
\boldsymbol{A}^{\prime} {\boldsymbol{x}^\prime _{\text{r}}} \leq \boldsymbol{b}^{\prime} \text{,} \ \ \ f^{\prime}>0  \text{.}  \notag
\end{align}
The objective function is not jointly convex in $f^{\prime}$ and ${\boldsymbol{x}^\prime _{\text{r}}}$, as shown for the conventional MMSE cost function in the Appendix \ref{app:Hessian_MMSE}.

However, it is possible to shift the 
scaling factor from the objective in the polyhedron constraint as done in Section~\ref{subsec:Proposed_Mapped_Precoder}.
Accordingly, we substitute the variable with $f^{\prime} {\boldsymbol{x}^\prime _{\text{r}}} =\boldsymbol{x}^{\prime}_{\text{r}, f}$.
Using $\boldsymbol{x}^{\prime}_{\text{r}, f}$, the equivalent problem reads as
\begin{align}
\label{eq:partial_problem2}
& \min_{\boldsymbol{x}^{\prime}_{\text{r}, f},f^{\prime}}  
 \lVert  
\boldsymbol{H}_{\text{r}}^{\prime} \ \boldsymbol{x}^{\prime}_{\text{r}, f} 
  -  \boldsymbol{s}_{\text{r}} +  f^{\prime}  \boldsymbol{H}_{\text{r, fixed}}\
\boldsymbol{x}_{\text{r, fixed}}        \rVert_2^2  +  f^{\prime}{}^2 
   \mathrm{E} \{  \boldsymbol{w}_{\text{r}}^T
   \boldsymbol{w}_{\text{r}}
  \} 
 \\
& \text{subject to: }     
\boldsymbol{A}^{\prime} \boldsymbol{x}^{\prime}_{\text{r}, f} \leq  f^{\prime}  \boldsymbol{b}^{\prime} \text{,} \ \ \ f^{\prime}>0  \text{.}  \notag
\end{align}
Rearranging the constraint yields
\begin{align}
\label{eq:partial_problem_final}
& \min_{\boldsymbol{x}^{\prime}_{\text{r}, f},f^{\prime}}  
 \lVert  
\boldsymbol{H}_{\text{r}}^{\prime}\ \boldsymbol{x}^{\prime}_{\text{r}, f}
  -  \boldsymbol{s}_{\text{r}} +  f^{\prime}  \boldsymbol{H}_{\text{r, fixed}}
\ \boldsymbol{x}_{\text{r, fixed}}        \rVert_2^2  +  f^{\prime}{}^2 
   \mathrm{E} \{  \boldsymbol{w}_{\text{r}}^T
   \boldsymbol{w}_{\text{r}}
  \} 
 \\
& \text{subject to: }     
 \boldsymbol{R}^{\prime}
 \begin{bmatrix}
 \boldsymbol{x}^{\prime}_{\text{r}, f}\\
f^{\prime}
\end{bmatrix}
  \leq  \boldsymbol{0}\text{,} \ \ \ f^{\prime}>0   \text{,}  \notag
\end{align}
where $\boldsymbol{R}^{\prime}=\pr{ \boldsymbol{A}^{\prime}  -\boldsymbol{b}^{\prime}  }$ is obtained by selecting the last $2\pc{M-d}$ columns of $\boldsymbol{R}$. 

Note that the problem in \eqref{eq:partial_problem_final} is convex because of the convex constraints and its objective function which is jointly convex in $f^{\prime}$ and $\boldsymbol{x}^{\prime}_{\text{r}, f}$ as can be seen in the Appendix \ref{app:Partial_MMSE}, where the Hessian is examined.
\paragraph{MMSE Branch-and-Bound Precoding Algorithm}
\label{sec:bb_algorithm_design}
$\\$
In this subsection, a branch-and-bound algorithm is proposed which solves \eqref{LRA_MMSE_precoder} with the tools presented in the previous subsections. As mentioned before, the first step is the initialization, where the problem from \eqref{eq:mapped_mmse_problem} is solved and the condition $\boldsymbol{x}_{\textrm{lb}}=\boldsymbol{x}_{\textrm{ub}}$ is evaluated. If the condition is met, the algorithm returns $\boldsymbol{x}_\textrm{lb}$. Otherwise, the branch-and-bound tree search process is performed as described in the sequel.

\begin{algorithm}
\small
  \caption{Proposed B\&B Precoding for solving \eqref{LRA_MMSE_precoder}}
	\label{alg:BB_Precoding}
  \begin{algorithmic}    
	\State{\textcolor{r}{Given the channel $\boldsymbol{H}$ and transmit symbols $\boldsymbol{s}$ compute a valid upper bound $\check{g}(\check{\boldsymbol{x}})$ on the problem in \eqref{LRA_MMSE_precoder}, by solving \eqref{eq:mapped_mmse_problem} followed by mapping to the closest precoding vector $\check{\boldsymbol{x}} \in \mathcal{X}_{\textrm{}}^{M}$, and computing its MSE. If the solution of \eqref{eq:mapped_mmse_problem} belongs to $\mathcal{X}_{\textrm{}}^{M}$ it is optimal. Otherwise, define the first level ($d=1$) of the tree by $\mathcal{G}_{d}:=\mathcal{X}_{\textrm{}}$}}
	\For{$d=1:M-1$}
	\State{ Partition  $\mathcal{G}_{d}$ in $\boldsymbol{x}_{\textrm{fixed},1},\ldots,\boldsymbol{x}_{\textrm{fixed},\left|\mathcal{G}_{d}\right|}$ }  
	  \For{$i=1:\left| \mathcal{G}_{d} \right|$}
		\State{Express $\boldsymbol{x}_{\textrm{fixed},i}$ in real valued notation $\boldsymbol{x}_{ \textrm{r,fixed},i}$ }
		\State{Conditioned on $\boldsymbol{x}_{\text{r} \textrm{,fixed},i}$ solve \eqref{eq:partial_problem_final} to find $\boldsymbol{x}^\prime_{\textrm{r},f}$ and $f^\prime$}
		\State{\textcolor{r}{Determine the lower bound as  $		\textrm{MSE}_\textrm{lb}:=		    		\PM{ \PM{
        \boldsymbol{H}_{\text{r}}^{\prime}\ \boldsymbol{x}^{\prime}_{\text{r}, f}
         -\boldsymbol{s}_{\text{r}} +  f^{\prime}  \boldsymbol{H}_{\text{r, fixed}}\
        \boldsymbol{x}_{\text{r, fixed,i}} } }_2^2 + {f^{\prime}}^2 
        \mathrm{E} \{  \boldsymbol{w}_{\text{r}}^T
        \boldsymbol{w}_{\text{r}}\}$}}
		\State{\textcolor{r}{Extract $\boldsymbol{x}_\text{r}^\prime= \frac{\boldsymbol{x}_{r,f}^\prime}{f^\prime}$ and rewrite $\boldsymbol{x}_\text{r}^\prime$ in complex notation as $\boldsymbol{x}_{\text{lb}}^\prime$}}
		\State{Map $\boldsymbol{x}_{\text{lb}}^\prime$ to the discrete solution with the closest Euclidean distance: ${\boldsymbol{x}}_\textrm{ub}^\prime(\boldsymbol{x}_{\mathrm{lb}}^\prime) \in \mathcal{X}^{M-d} $} 
	  \State{Express $\boldsymbol{x}_\textrm{ub}^\prime$ in real valued notation $\boldsymbol{x}_{\textrm{r,ub}}^\prime$ and compute $f^\prime$ according to \eqref{eq:optimal_f_prime}}
		\State{\textcolor{r}{With $\boldsymbol{x}_{\textrm{r,ub}}^\prime$ and $f^\prime$, the upper bound is $\textrm{MSE}_\textrm{ub}(\boldsymbol{x}_{\textrm{ r,fixed,i}}):=
		\PM{\PM{\
        f^\prime \boldsymbol{H}_{\text{r}} 
        \begin{bmatrix}
        \boldsymbol{x}_{\textrm{r,fixed,i}} \ \
        \boldsymbol{x}^\prime_{\text{r,ub}} 
        \end{bmatrix} 
         -\boldsymbol{s}_{\text{r}}}}_2^2  
         +
         f^{\prime}{}^2 
        \mathrm{E} \{  \boldsymbol{w}_{\text{r}}^T
        \boldsymbol{w}_{\text{r}}
        \}$
		}}
		\State{Update the best upper bound with $\check{g} =\min\left( \check{g}, \textrm{MSE}_\textrm{ub}  \right)$ and update $\check{\boldsymbol{x}}$ accordingly} 
	\EndFor
	\State{Construct a reduced set by comparing conditioned  lower bounds with the global upper bound $\check{g}$}
	\State{$\mathcal{G}_{d}^{\prime}:=\left\{  \boldsymbol{x}_{\textrm{lb},i}^\prime \vert \textrm{MSE}_\textrm{lb}(\boldsymbol{x}_{\textrm{lb},i}^\prime)  <  \check{g}       , i=1,\ldots,  \left|\mathcal{G}_{d}\right|  \right\} $}
	\State{\textcolor{r}{Define the set for the next level in the tree: $\mathcal{G}_{d+1}:=\mathcal{G}_{d}^{\prime} \times \mathcal{X}_{\textrm{}}$}}
	\EndFor
	\vspace{1mm}
	\State{Search method for the ultimate level $d=M$,}
	\vspace{1mm}
	\State{Partition $\mathcal{G}_{M}$ in $\boldsymbol{x}_{\textrm{fixed},1},\ldots,\boldsymbol{x}_{\textrm{fixed},\left|\mathcal{G}_{M}\right|}$}
	\State{Express $\boldsymbol{x}_{\textrm{fixed},i}$ with real valued notation $\boldsymbol{x}_{\textrm{r,fixed},i}$ and compute $f^\prime_i$ with \eqref{eq:optimal_f_prime}}
	  \begin{align*}
		\textrm{MSE}(\boldsymbol{x}&_{\textrm{fixed},i}) :=		
		\PM{\PM{
        f_i^\prime \boldsymbol{H}_{\text{r}} 
         \boldsymbol{x}_{\textrm{r,fixed,i}}
         -\boldsymbol{s}_{\text{r}}}}_2^2  
         +
         f_i^{\prime}{}^2 
        \mathrm{E} \{  \boldsymbol{w}_{\text{r}}^T
        \boldsymbol{w}_{\text{r}}
        \} 
		\end{align*}	
\State{The global solution is $\boldsymbol{x}_{\textrm{opt}} = \mathrm{arg} \displaystyle\hspace{-1.2em} \min_{\boldsymbol{x} \in \mathcal{G}_{M} \cup \chav{\boldsymbol{\check{x}}}} \ \textrm{MSE}(\boldsymbol{x})$}
\end{algorithmic}
\end{algorithm}

For the tree search process a breadth first search is devised and the subproblems are solved considering partially fixed precoding vectors $\boldsymbol{x}_{\text{r}}= \pr{ \boldsymbol{x}_{\text{r,fixed}}^T \  \ {\boldsymbol{x}^\prime_{\text{r}}}^{T}}^T$, where $\boldsymbol{x}_{\text{r,fixed}}$ has length $2d$ as previously stated, \textcolor{r}{where $d$ represents the layer of the tree as shown in Fig.~\ref{fig:tree}}. Accordingly, the matrix $\boldsymbol{H}_\textrm{r}$ is divided as $\boldsymbol{H}_\textrm{r}=\pr{\boldsymbol{H}_{\textrm{r},\textrm{fixed}} \ \ \boldsymbol{H}_\textrm{r}^{\prime}}$, where $\boldsymbol{H}_{\textrm{r},\textrm{fixed}}$ contains the first $2d$ columns of $\boldsymbol{H}_\textrm{r}$. Moreover the matrix $\boldsymbol{R}^\prime$ is obtained via selecting the last $2\pc{M-d}$ columns of $\boldsymbol{R}$.

Using $\boldsymbol{R}^\prime$ and $\boldsymbol{H}_\textrm{r}^\prime$, the subproblem \eqref{eq:partial_problem_final} for the lower-bounding step is solved. Mapping the solution from \eqref{eq:partial_problem_final} to the discrete set yields $\boldsymbol{x}_{\text{r,ub}}$. Based on $\boldsymbol{x}_{\text{r,ub}}$, the MSE is minimized by
\begin{align}
    \label{eq:optimal_f_prime}
    f^{\prime}=	
    \frac{\boldsymbol{s}_\textrm{r}^T 
    \boldsymbol{x}_{\boldsymbol{H_\text{r},\text{fixed}}}}
{ \PM{\PM{\boldsymbol{x}_{\boldsymbol{H_\text{r},\text{fixed}}}}}_2^2+        \mathrm{E} \{  \boldsymbol{w}_{\text{r}}^T
        \boldsymbol{w}_{\text{r}} 
        \} 
          }\textrm{.}
\end{align}
where $\boldsymbol{x}_{\boldsymbol{H_\text{r},\text{fixed}}}=\boldsymbol{H}_{\text{r}} \pr{\boldsymbol{x}_{\text{r,fixed}}^T \ \boldsymbol{x}_{\text{r,ub}}^T}^T$.
The corresponding MSE serves as an upper bound on the optimal value of the original problem ($\textrm{MSE}_\textrm{ub}$).
In case the lower bound conditioned on $\boldsymbol{x}_{\text{r,fixed}}$ is higher than any upper bound on the original problem, $\boldsymbol{x}_{\text{r,fixed}}$ cannot be part of the solution and every member of the discrete solution set which includes $\boldsymbol{x}_{\text{r,fixed}}$ can be excluded from the search process.
The steps of the method are detailed in Algorithm~\ref{alg:BB_Precoding}. 

Note that, when operating in the high-SNR regime, the computation of the optimal precoding vector in each symbol period can correspond to enormous computational complexity. A precomputation of the lookup-table $\mathcal{L}$ can be relevant, since it allows the precoding method to be a practical solution for channels with large coherence time, as suggested in \cite{Jedda_2016}.

\section{Receiver Design}
\label{sec:receiver}

This section exposes the design of the DPA-IDD Receiver where three soft detection methods for the computation of the extrinsic information are proposed. The objective of the DPA-IDD receiver is to compute LLRs that are used to make a decision about ${\boldsymbol{c}_k}$ which implies ${\boldsymbol{m}_k}$.
The LLRs are defined as follows 
\begin{equation}
\label{eq:L1}
L({c}_{k,i})=\ln\pc{\frac{
{P}\pc{{c}_{k,i} = +1 | {z}_k[t]}}
{{P}\pc{{c}_{k,i} = -1 | {z}_k[t]}}}\text{,}
\end{equation}
where ${z}_k[t]$ is the received signal and  ${c}_{k,i} \in \{-1,+1\}$. Using Bayes' theorem \eqref{eq:L1} is rewritten as \looseness-1
\begin{align}
\label{eq:a_posteriori_LLR}
L({c}_{k,i})&=\ln\pc{\frac{
{p}\pc{{z}_k[t]|{c}_{k,i} = +1}}
{{p}\pc{{z}_k[t]|{c}_{k,i} = -1}}
}+
\ln\pc{\frac{
{P}\pc{{c}_{k,i} = +1}}
{{P}\pc{{c}_{k,i} = -1}}
}\notag\\[5pt]
&=L_e \pc{c_{k,i}} + L_a\pc{{c}_{k,i}} ,
\end{align}
where $L_e \pc{c_{k,i}}$ and $L_a \pc{c_{k,i}}$ denote the extrinsic and a priori information, respectively. 

\subsection{Extrinsic Information Computation}
\label{subsec:extrinsic_llr}

As shown in equation \eqref{eq:a_posteriori_LLR}, the $L_e(c_{k,i})$ is defined as 
\begin{align}
\label{eq:extrinsic_LLR_definition}
    L_e \pc{c_{k,i}}=\ln\pc{\frac{
{p}\pc{{z}_k[t]|{c}_{k,i} = +1}}
{{p}\pc{{z}_k[t]|{c}_{k,i} = -1}}
}.
\end{align}
Using the law of total probability equation \eqref{eq:extrinsic_LLR_definition} can be expanded as
\begin{align}
\label{eq:extrinsic_LLR}
L_e \pc{c_{k,i}}&=
\ln\pc{\frac{
\displaystyle \sum_{s\in S_{+1}}{p}\pc{{z}_k[t]|s} {P}\pc{s|{r}_{k,t,\upsilon} = +1}}
{\displaystyle \sum_{s\in S_{-1}}{p}\pc{{z}_k[t]|s}{P}\pc{s|{r}_{k,t,\upsilon} = -1}}
}\text{,}
\end{align}
with $\upsilon=\pc{i-(t-1)N} \in \{1,\hdots, N\}$. The sets $S_{+1}$ and $S_{-1}$ represent all possible constellation points where the $\upsilon$-th bit of $\boldsymbol{r}_{k,t}$ is $+1$ or $-1$, respectively.
For a given $s\in S_g$, $g \in \{+1,-1\}$, if $\mathcal{M}^{-1}(s)=[a_{1},\hdots, a_{\upsilon}=g, \hdots,a_{N}]$, the probability ${P}\pc{s|{r}_{k,t,\upsilon}=g}$ is given by
\begin{align}
    \label{eq:probabilities}
    {P}\pc{s|{r}_{k,t,\upsilon}=g}=\displaystyle\prod_{\substack{l=1 \\[2pt] l\neq \upsilon}}^{N} P(r_{k,t,l}=a_{l}).
\end{align}
Based on the a priori information, the previous equation is rewritten as
\begin{align}
    \label{eq:a_priori_llr}
    P\pc{s|r_{k,t,\upsilon}=g}=\displaystyle\prod_{\substack{l=1 \\[2pt] l\neq \upsilon}}^N {
    \frac
    {e^{\pc{a_l \ L_a(r_{k,t,l})}}} 
    {1+e^{\pc{a_l \ L_a(r_{k,t,l})}}}
    } ,
\end{align}
where $L_a\pc{{r}_{k,t,l}}=L_a\pc{{c}_{k,l+(t-1)N}}$. Note that for computing $L_e(c_{k,i})$ the only demands are $p\pc{{z}_k[t]|s}$, that need to be known for all $s\in \mathcal{S}$, and the knowledge of $L_a\pc{c_{k,i}}$ for $i=1,\hdots, \frac{N_b}{R}$.

\subsubsection{Discrete Precoding Aware Soft Detector}
\label{subsec:optimal_llr}
$\\$
In this subsection, we introduce the DPA Soft Detector as a method for computing $L_e(c_{k,i})$. 
First, the received signal $z_k[t]$ is rewritten in a stacked vector notation 
$ \boldsymbol{z}_r[t] = [ \text{Re}\chav{z_k[t]} \  \text{Im}\chav{z_k[t]} ]^T$, where, for simplicity, the index $k$ is suppressed. 
The distribution $p\pc{{z}_k[t]|s}$ is given by
\begin{align}
    \label{eq:real_pdf}
    p\pc{{z}_k[t]|s}&=\displaystyle\sum_{\boldsymbol{s^\prime}\in \mathcal{S}^{K-1}} p\pc{{z}_k[t]|s,\boldsymbol{s}^\prime}\ P(\boldsymbol{s}^\prime)\notag\\
    &=\pc{\frac{1}{\alpha_s}}^{K-1} \frac{1}{\pi \sigma_w^2}\displaystyle\sum_{\boldsymbol{s^\prime}\in \mathcal{S}^{K-1}}
    \text{e}^{-\frac{\PM{\PM{\boldsymbol{z}_r[t]-\text{E}\chav{\boldsymbol{z}_r[t]|s,\boldsymbol{s}^\prime}}}_2^2}{\sigma_w^2}},
\end{align}
where $\boldsymbol{s}^\prime=\pr{s_1^\prime,\hdots,s_{k-1}^\prime,s_{k+1}^\prime,\hdots,s^\prime_K}^T$ corresponds to the symbols of the other users. For a given $s$ and $\boldsymbol{s}^\prime$ the expected value of the receive signal is given by
\begin{align}
    \text{E}\chav{\boldsymbol{z}_r[t]|s,\boldsymbol{s}^\prime}=[\text{Re}\chav{\boldsymbol{h}_k\boldsymbol{x}\pc{s,\boldsymbol{s}^\prime}} \ \ \text{Im}\chav{\boldsymbol{h}_k\boldsymbol{x}\pc{s,\boldsymbol{s}^\prime}}]^T.
\end{align}
With this, $L_e(c_{k,i})$ can be computed by inserting \eqref{eq:real_pdf} into \eqref{eq:extrinsic_LLR}. The resulting expression, finally, reads as
\begin{align}
\label{eq:optimal_llr}
    L_e(c_{k,i})&=\ln\pc{\frac{
\displaystyle \sum_{s\in S_{+1}}
\displaystyle\sum_{\boldsymbol{s^\prime}\in \mathcal{S}^{K-1}}
    \text{e}^{-\frac{\PM{\PM{\boldsymbol{z}_r[t]-\text{E}\chav{\boldsymbol{z}_r[t]|s,\boldsymbol{s}^\prime}}}_2^2}{\sigma_w^2}}
{P}\pc{s|{r}_{k,t,\upsilon} = +1}}
{\displaystyle \sum_{s\in S_{-1}}
\displaystyle\sum_{\boldsymbol{s^\prime}\in \mathcal{S}^{K-1}}
    \text{e}^{-\frac{\PM{\PM{\boldsymbol{z}_r[t]-\text{E}\chav{\boldsymbol{z}_r[t]|s,\boldsymbol{s}^\prime}}}_2^2}{\sigma_w^2}}
{P}\pc{s|{r}_{k,t,\upsilon} = -1}}
}\text{.}
\end{align}
Note that, for using \eqref{eq:optimal_llr}, $p\pc{{z}_k[t]|s,\boldsymbol{s}^\prime}$ needs to be evaluated for all members of $\mathcal{S}^{K-1}$. Hence, computing \eqref{eq:optimal_llr} can lead to a prohibitive computational complexity at the receiver side for systems with many users. 

\subsubsection{Gaussian Discrete Precoding Aware Soft Detector}
\label{subsec:exact_llr}
$\\$
To reduce the computational complexity we introduce the Gaussian Discrete Precoding Aware (GDPA) Soft Detector. 
The basic assumption is that the vector $\boldsymbol{z}_r[t]$ can be described as a Gaussian random vector, meaning
\begin{align}
\label{eq:channel_probability}
    \tilde{p}\pc{{z}_k[t] |   s}=\frac{ \text{e}^{-\frac{1}{2} 
   \pr{\pc{\boldsymbol{z}_r[t]-\boldsymbol{\mu}_{z_r|s} }^T\boldsymbol{C}_{z_r|s}^{-1}\pc{\boldsymbol{z}_r[t]-\boldsymbol{\mu}_{z_r|s}}}} }{2\pi \sqrt{\text{det}\pc{\boldsymbol{C}_{z_r|s}}}}.
\end{align}
In the following the computation of $\boldsymbol{\mu}_{z_r|s}$ and $\boldsymbol{C}_{z_r|s}$ is detailed. Since $\text{E}\chav{\text{Re}\chav{a}}=\text{Re}\chav{\text{E}\chav{a}}$, and $\text{E}\chav{\text{Im}\chav{a}}=\text{Im}\chav{\text{E}\chav{a}}$, we compute first the expected value of the complex received signal, which reads as
\begin{align}
\label{eq:mean}
\text{E}\chav{z_k[t]|  s}
&=\text{E}\chav{{ {\boldsymbol{h}_k\boldsymbol{x}[t]}}\ |  s}.
\end{align}
In order to simplify the notation we introduce the variable $\zeta(\boldsymbol{s})=\boldsymbol{h}_k\ \boldsymbol{x}\pc{\boldsymbol{s}}$. The mean vector $\boldsymbol{\mu}_{z_r|s}$ is, then, given by
\begin{align}
\label{eq:mean_vector}
\boldsymbol{\mu}_{z_r|s}&=\begin{bmatrix} 
    \text{E}\chav{\text{Re}\chav{\zeta\pc{\boldsymbol{s}}} |  s}  \ \ \ 
    \text{E}\chav{\text{Im}\chav{\zeta\pc{\boldsymbol{s}}} |  s}
\end{bmatrix} ^T,
\end{align}
where
\begin{align}
\label{eq:expected_values1}
\text{E}\chav{\text{Re}\chav{\zeta\pc{\boldsymbol{s}}}|{s}}&=\pc{\frac{1}{\alpha_s}}^{K-1}\displaystyle \sum_{\boldsymbol{s}\in \mathcal{D}}{ \text{Re}\chav{\zeta\pc{\boldsymbol{s}}}},\\
\label{eq:expected_values2}
\text{E}\chav{\text{Im}\chav{\zeta\pc{\boldsymbol{s}}}|{s}}&=\pc{\frac{1}{\alpha_s}}^{K-1}\displaystyle \sum_{\boldsymbol{s}\in \mathcal{D}}{\text{Im}\chav{\zeta\pc{\boldsymbol{s}}}}
\end{align}
and $\mathcal{D}$ is the set of all possible $\boldsymbol{s}[t]$ whose k-th entry is $s$. Moreover, the corresponding covariance matrix is given by
\begin{align}
\label{eq:czr|s}
    \boldsymbol{C}_{z_r|s}= \begin{bmatrix} 
    \sigma_{r|s}^2\ \ \ &\rho_{ri|s}  \\  
    \rho_{ri|s} \ \ \ &\sigma_{i|s}^2 
\end{bmatrix}.
\end{align}
The entries of $\boldsymbol{C}_{z_r|s}$ reads as
 \begin{align}
      \sigma_{r|s}^2&=\frac{\sigma_w^2}{2}+ \text{E}\chav{\text{Re}\chav{\zeta\pc{\boldsymbol{s}}}^2|{s}}- \text{E}\chav{\text{Re}\chav{\zeta\pc{\boldsymbol{s}}}|{s}}^2,  \\[5pt]
      \sigma_{i|s}^2&=\frac{\sigma_w^2}{2}+\text{E}\chav{\text{Im}\chav{\zeta\pc{\boldsymbol{s}}}^2|{s}}- \text{E}\chav{\text{Im}\chav{\zeta\pc{\boldsymbol{s}}}|{s}}^2, \\[5pt]
     \rho_{ri|s}&=\text{E}\{\text{Re}\{\zeta\pc{\boldsymbol{s}}\}\text{Im}\{\zeta\pc{\boldsymbol{s}}|s\}\}-  \text{E}\{\text{Re}\{\zeta\pc{\boldsymbol{s}}\}|s\}\text{E}\{\text{Im}\chav{\zeta\pc{\boldsymbol{s}}}|s\},
 \end{align}
where 
\begin{align}
\text{E}&\chav{\text{Re}\chav{\zeta\pc{\boldsymbol{s}}}^2|{s}}=\pc{\frac{1}{\alpha_s}}^{K-1}\displaystyle \sum_{\boldsymbol{s}\in \mathcal{D}}{\text{Re}\chav{\zeta\pc{\boldsymbol{s}}}^2},\\
\text{E}&\chav{\text{Im}\chav{\zeta\pc{\boldsymbol{s}}}^2|{s}}=\pc{\frac{1}{\alpha_s}}^{K-1}\displaystyle \sum_{\boldsymbol{s}\in \mathcal{D}}{\text{Im}\chav{\zeta\pc{\boldsymbol{s}}}^2},\\
\text{E}&\chav{\text{Re}\chav{\zeta\pc{\boldsymbol{s}}}\text{Im}\chav{\zeta\pc{\boldsymbol{s}}}|s}=
\pc{\frac{1}{\alpha_s}}^{K-1}\displaystyle \sum_{\boldsymbol{s}\in \mathcal{D}}{\text{Re}\chav{\zeta\pc{\boldsymbol{s}}}\text{Im}\chav{\zeta\pc{\boldsymbol{s}}}}
\end{align} 
and $\text{E}\chav{\text{Re}\chav{\zeta\pc{\boldsymbol{s}}}|s}$ and $\text{E}\chav{\text{Im}\chav{\zeta\pc{\boldsymbol{s}}}|s}$ are defined in equations \eqref{eq:expected_values1} and \eqref{eq:expected_values2}, respectively. Based on $\boldsymbol{C}_{z_r|s}$ and $\boldsymbol{\mu}_{z_r|s}$,  $L_e(c_{k,i})$ is computed as
\begin{align}
\label{eq:LLR_GDPA}
    L_e(c_{k,i})&=\ln\pc{\frac{
\displaystyle \sum_{s\in S_{+1}}
\frac{\text{e}^{\Psi_s} }
{\sqrt{\text{det}\pc{{\boldsymbol{C}_{z_r|s}}}}}
{P}\pc{s|{r}_{k,t,\upsilon} = +1}}
{\displaystyle \sum_{s\in S_{-1}}
\frac{\text{e}^{\Psi_s} }
{\sqrt{\text{det}\pc{{\boldsymbol{C}_{z_r|s}}}}}
{P}\pc{s|{r}_{k,t,\upsilon} = -1}}
}\text{,}
\end{align}
where
\begin{align}
    \Psi_{s}=-\frac{1}{2} 
   \pr{\pc{\boldsymbol{z}_r[t]-\boldsymbol{\mu}_{z_r|s} }^T\boldsymbol{C}_{z_r|s}^{-1}\pc{\boldsymbol{z}_r[t]-\boldsymbol{\mu}_{z_r|s}}}
\end{align}
and ${P}\pc{s|{r}_{k,t,\upsilon} = g}$ for $g\in\{-1,+1\}$ can be computed with equation \eqref{eq:a_priori_llr} considering $L_a(c_{k,i})$ for $i=1,\hdots,\frac{N_b}{R}$.

Note that, when calculating $L_e(c_{k,i})$ using \eqref{eq:LLR_GDPA}, $\tilde{p}({z_k}[t]|s)$ is evaluated only $\alpha_s$ times. This results in a significant decrease in computational complexity, when compared with the approach proposed in \eqref{eq:optimal_llr}. 
However, for computing \eqref{eq:LLR_GDPA}, the receiver requires access to $\boldsymbol{C}_{z_r|s}$ and $\boldsymbol{\mu}_{z_r|s}$ for all values of $s$. These parameters need to be provided by the BS which causes communication overhead. In this context, an alternative method that requires a fewer number of parameters to be transmitted is desired.

\subsubsection{Linear Model Based Discrete Precoding Aware Soft Detector}
\label{subsec:Linear_Model}
$\\$
In this subsection, a method for computing $L_e(c_{k,i})$ with a reduced number of model parameters is devised. This proposed approach relies on the description of the received signal by a linear model.\looseness-1

\paragraph{Discrete Precoding Aware Linear Model}
\label{subsec:introduction_on_weighting}
$\\$
The Discrete Precoding Aware Linear Model (DPA-LM) is based on the assumption that the received signal can be expressed by 
\begin{align}
    \label{eq:channel_model}
        {z}_k[t]=h_{k}^\text{eff}{s_k[t]+w_k[t]+\epsilon_k[t]} \text{,}
\end{align}
where $h^\text{eff}_k \in \mathcal{C}$ is a factor that expresses the precoder and channel effects on the transmit symbol of the $k$-th user and $\epsilon_k[t]$ is the error term that denotes the difference between $z_k[t]$ and $h_{k}^\text{eff}s_k[t]+w_k[t]$. To identify an appropriate $h^\text{eff}_k$ we consider the following MSE optimization problem \looseness-1
 \begin{align}
 \label{eq:problem_for_heff}
     h^\text{eff}_k&=\arg\min \lambda_{\epsilon_k}^2 =\arg\min \text{E}\chav{ \PM{ \epsilon_k[t]    }^2}\notag\\
     &= \arg\min_{\gamma \in \mathcal{C  }}\text{E}\chav{ \PM{ \boldsymbol{h}_k \ \boldsymbol{x}[t] - {\gamma \ s_k[t]}}^2}\text{,}
 \end{align}
where the optimal solution is given by
 \begin{align}
     h^\text{eff}_k&={\frac{1}{\alpha_s^K\ \sigma_s^2}}\sum_{\boldsymbol{s} \in \mathcal{S}^K} \ s_k^*(\boldsymbol{s}) \ \zeta(\boldsymbol{s}),\\
    \lambda_{\epsilon_k}^2&=
    \boldsymbol{h}_k\ \boldsymbol{\Lambda}_x \ \boldsymbol{h}^H_k-\PM{h^\text{eff}_k}^2 \sigma_s^2, 
 \end{align}
where $\boldsymbol{\Lambda}_x=\pc{\frac{1}{\alpha_s}}^K \displaystyle \sum_{\boldsymbol{s}\in \mathcal{S}^K} {\boldsymbol{x}\pc{\boldsymbol{s}} \boldsymbol{x}\pc{\boldsymbol{s}}^H}$ and $s_k\pc{\boldsymbol{s}}$ is the $k$-th element of $\boldsymbol{s}$. The derivation of the values for $h^\text{eff}_k$ and $\lambda_{\epsilon_k}^2$ is given in the Appendix \ref{subsec:derivation_of_heff}.

\paragraph{DPA-LM Soft Detector}
\label{subsec:LLR_approximation}
$\\$
This subsection describes the proposed DPA-LM Soft Detector as a method for computing the extrinsic information based on the linear model previously presented.

The following strategy relies on the assumption that the error term $\epsilon_k[t]$ is a circular symmetric complex Gaussian random variable. 
The expected value of the received signal is calculated as
\begin{align}
\text{E}\chav{z_k[t]|s}&= h_k^\text{eff} s +  \text{E}\chav{\epsilon_k[t]|s}, 
\end{align}
and assuming $\text{E}\chav{\epsilon_k[t]|s}=0 \ \forall \ s \in \mathcal{S}$ yields
\begin{align}    
\label{eq:muzrs}
\boldsymbol{\mu}_{z_r|s}^{\text{eff}}&= \begin{bmatrix} 
    \text{Re}\chav{h_k^\text{eff} s}  \ \ \  
    \text{Im}\chav{h_k^\text{eff} s}
\end{bmatrix}^T . 
\end{align}
Considering that 
\begin{align}
\label{eq:czr_s}
    \boldsymbol{C}_{z_r}^{\text{eff}}=\frac{{\sigma}_{\text{eff}_k}^2}{2} \boldsymbol{I},
\end{align}
with ${\sigma}_{\text{eff}_k}^2={\lambda}_{\epsilon_k}^2+ \sigma_w^2$ being the effective noise variance. Then, the extrinsic information function from \eqref{eq:extrinsic_LLR} simplifies to 
\begin{align}
\label{eq:linear_model_LLR}
   \hspace{-1.885mm}
   L_e\pc{c_{k,i}}=
    \text{ln}\pc{
    \frac{\displaystyle \sum_{s\in S_{+1}}  \text{e}^{-\frac{\PM{{z_k[t]-h_k^\text{eff}\ s}}^2}{{\sigma}_{\text{eff}_k}^2} } \ {P}\pc{s|{r}_{k,t,\upsilon} = +1}
}
    {\displaystyle\sum_{s\in S_{-1}}  \text{e}^{-\frac{\PM{{z_k[t]-h_k^\text{eff}\ s}}^2}{{\sigma}_{\text{eff}_k}^2}}\ {P}\pc{s|{r}_{k,t,\upsilon} = -1}}
    } ,
\end{align}
where the values for ${P}\pc{s|{r}_{k,t,\upsilon} = g}$, for $g \in \{-1,+1\}$ are calculated using equation \eqref{eq:a_priori_llr}.

The computation of $L_e(c_{k,i})$ according to \eqref{eq:linear_model_LLR} only requires knowledge about the parameters $h_k^\text{eff}$ and ${\sigma}_{\text{eff}_k}^2$, which are independent of the data symbol $s$. In comparison with the method from subsection~\ref{subsec:exact_llr}, the number of parameters that need to be transmitted in advance to the information data is significantly reduced.

\subsection{DPA-IDD Scheme}
\label{subsec:dpa_idd_algorithm}

Subsections \ref{subsec:optimal_llr}, \ref{subsec:exact_llr} and \ref{subsec:LLR_approximation} expose different methods for computing $L_e(c_{k,i})$ when $L_a(c_{k,i})$ is known. Using these results, the DPA-IDD scheme is presented as a way of computing $L(c_{k,i})$ via making an iterative estimation of $L_a(c_{k,i})$ and, consequently, $L_e(c_{k,i})$. For description of the DPA-IDD scheme, we define 
\begin{align}
\hspace{-4mm}
\quad
\boldsymbol{L}&=
\begin{bmatrix}
   L(c_{k,1})   
   \hdots
   L(c_{k,\frac{N_b}{R}})
\end{bmatrix},
\quad
\boldsymbol{L}_e=\begin{bmatrix}
   L_e(c_{k,1}) 
   \hdots 
   L_e(c_{k,\frac{N_b}{R}})
\end{bmatrix},
\quad
\boldsymbol{L}_a=\begin{bmatrix}
   L_a(c_{k,1}) 
   \hdots
   L_a(c_{k,\frac{N_b}{R}})
   \end{bmatrix}.
   \notag
\end{align}
The principle of the proposed receiver is based on equation \eqref{eq:a_posteriori_LLR}. Based on $\boldsymbol{L}$ and $\boldsymbol{L}_e$, the a priori information is extracted via $\boldsymbol{L}_a=\boldsymbol{L}-\boldsymbol{L}_e$. 

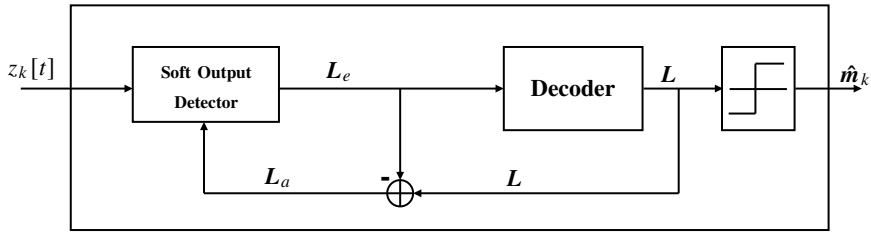
\begin{figure}[ht]
\begin{center}
\tikzset{every picture/.style={line width=0.75pt}} 

\begin{tikzpicture}[x=0.35pt,y=0.35pt,yscale=-1,xscale=1]

\draw   (630,189) -- (779.93,189) -- (779.93,279) -- (630,279) -- cycle ;
\draw    (108,234) -- (226,234) ;
\draw [shift={(229,234)}, rotate = 180] [fill={rgb, 255:red, 0; green, 0; blue, 0 }  ][line width=0.08]  [draw opacity=0] (8.93,-4.29) -- (0,0) -- (8.93,4.29) -- cycle    ;
\draw   (162,144) -- (981,144) -- (981,387) -- (162,387) -- cycle ;
\draw    (819,234) -- (819,347.23) ;
\draw    (535,347.23) -- (819,347.23) ;
\draw [shift={(532,347.23)}, rotate = 0] [fill={rgb, 255:red, 0; green, 0; blue, 0 }  ][line width=0.08]  [draw opacity=0] (8.93,-4.29) -- (0,0) -- (8.93,4.29) -- cycle    ;
\draw   (229,189) -- (387,189) -- (387,270) -- (229,270) -- cycle ;
\draw    (781,234) -- (864,234) ;
\draw [shift={(867,234)}, rotate = 180] [fill={rgb, 255:red, 0; green, 0; blue, 0 }  ][line width=0.08]  [draw opacity=0] (8.93,-4.29) -- (0,0) -- (8.93,4.29) -- cycle    ;
\draw    (387,234) -- (627,234) ;
\draw [shift={(630,234)}, rotate = 180] [fill={rgb, 255:red, 0; green, 0; blue, 0 }  ][line width=0.08]  [draw opacity=0] (8.93,-4.29) -- (0,0) -- (8.93,4.29) -- cycle    ;
\draw    (306,273) -- (306,347.23) ;
\draw [shift={(306,270)}, rotate = 90] [fill={rgb, 255:red, 0; green, 0; blue, 0 }  ][line width=0.08]  [draw opacity=0] (8.93,-4.29) -- (0,0) -- (8.93,4.29) -- cycle    ;
\draw    (306,347.23) -- (504,347.23) ;
\draw   (865.8,189) -- (944.24,189) -- (944.24,279) -- (865.8,279) -- cycle ;
\draw    (874.03,234.53) -- (936,234.53) ;
\draw    (901.92,207) -- (932.9,207) ;
\draw    (873,261) -- (901.92,261) ;
\draw    (901.92,207.13) -- (901.92,261) ;
\draw    (945,234) -- (1014,234) ;
\draw [shift={(1017,234)}, rotate = 180] [fill={rgb, 255:red, 0; green, 0; blue, 0 }  ][line width=0.08]  [draw opacity=0] (8.93,-4.29) -- (0,0) -- (8.93,4.29) -- cycle    ;
\draw   (504,347.23) .. controls (504,339.37) and (510.27,333) .. (518,333) .. controls (525.73,333) and (532,339.37) .. (532,347.23) .. controls (532,355.09) and (525.73,361.46) .. (518,361.46) .. controls (510.27,361.46) and (504,355.09) .. (504,347.23) -- cycle ; \draw   (504,347.23) -- (532,347.23) ; \draw   (518,333) -- (518,361.46) ;
\draw    (518,233.77) -- (518,330) ;
\draw [shift={(518,333)}, rotate = 270] [fill={rgb, 255:red, 0; green, 0; blue, 0 }  ][line width=0.08]  [draw opacity=0] (8.93,-4.29) -- (0,0) -- (8.93,4.29) -- cycle    ;

\draw (704.97,233.32) node  [scale=0.75] [align=center] {\textbf{${\text{Decoder}}$}};
\draw (120,212.5) node  [scale=0.75]  {$z_{k}[t]$};
\draw (808,218.5) node  [scale=0.75]  {$\boldsymbol{L}$};
\draw (641,330.5) node  [scale=0.75]  {$\boldsymbol{L}$};
\draw (308.64,232.21) node  [scale=0.55] [align=center] {\textbf{Soft Output}\\\textbf{Detector}};
\draw (385.5,330.5) node  [scale=0.75]  {$\boldsymbol{L}_a$};
\draw (451.5,215.5) node  [scale=0.75]  {$\boldsymbol{L}_e$};
\draw (1010,219.5) node  [scale=0.75]  {$\boldsymbol{\hat{m}}_k$};
\draw (502,335) node  [scale=1]  {\textbf{-}};

\end{tikzpicture}
\caption{DPA-IDD Receiver Topology} 
\label{fig:dpa_idd_topology}       
\end{center}
\end{figure}

With this, for initialization, the detector calculates $\boldsymbol{L}_e$ assuming $\boldsymbol{L}_a=\boldsymbol{0}$ and forwards it to the decoder. The decoder outputs the LLR vector $\boldsymbol{L}$. Using $\boldsymbol{L}$ and $\boldsymbol{L}_e$, the a priori information is calculated and fed back into the detector which will, then, recompute $\boldsymbol{L}_e$ based on the updated $\boldsymbol{L}_a$. This process is done recursively until the maximum number of iterations is reached. An illustration of the receiving process is shown in Fig.~\ref{fig:dpa_idd_topology}.

The DPA-IDD technique does not require a specific method for computing $\boldsymbol{L}_e$. Hence, the approaches presented in subsections \ref{subsec:optimal_llr}, \ref{subsec:exact_llr} and \ref{subsec:LLR_approximation} are compatible with the framework and can be used for calculating $\boldsymbol{L}_e$.

Note that, unlike other IDD approaches, e.g. \cite{Ten_brink_idd}, the proposed soft detectors compute $L_e(c_{k,i})$ instead of $L(c_{k,i})$. As a consequence, there is no need for subtracting the a priori information from the soft detector's output, as shown in Fig.~\ref{fig:dpa_idd_topology}.

\section{Numerical Results}
\label{sec:numerical_results}

For the numerical evaluation, the BER is considered. We assume that the channel gains are modeled by independent Rayleigh fading \cite{Marzetta_2013}, meaning $\beta_m=1 \ \text{for} \ m=1,...,M$ and $g_{k,m}\sim \mathcal{CN}  ({0},\sigma_g^2)\ \text{for} \ k=1,...,K \ \text{and}\ m=1, ...,M$ as done implicitly in \cite{ZF-Precoding} and \cite{MSM_precoder} and explicitly in \cite{CVX-CIO}, moreover, the SNR is defined by $\mathrm{SNR}=\frac{ \left\|\boldsymbol{x}\right\|^2_2 }{\sigma_w^2}$.

\subsection{Uncoded Transmission}

In this subsection the performance of the proposed precoding algorithms are evaluated against with following the state-of-the-art approaches:
\begin{itemize}
    \item[1.] The MSM-Precoder \cite{MSM_precoder} considering phase quantization;
    \item[2.] The ZF precoder with constant envelope \cite{ZF-Precoding}, where the entries of the precoding vector are subsequently phase quantized;
    \item[3.] The phase quantized CIO precoder implemented via CVX \cite{CVX-CIO};
    \item[4.] \textcolor{r}{The single-carrier version of the SQUID-OFDM precoder \cite{jacobsson2018nonlinear} phase quantized};
    \item[6.] \textcolor{r}{The C3PO precoder \cite{Struder_c3po} phase quantized considering a real valued scaling factor};
    \item[7.] The MMDDT branch-and-bound precoder \cite{General_MMDDT_BB};
\end{itemize}

In this context, the subsection is divided into two parts. In the first part, the precoding strategies are evaluated in terms of BER using phase quantizers as hard detectors. In the second part, the complexity of the proposed methods is analyzed and compared against the mentioned algorithms. 

\subsubsection{Hard Detection using Phase Quantizers at the Receiver}
$\\$
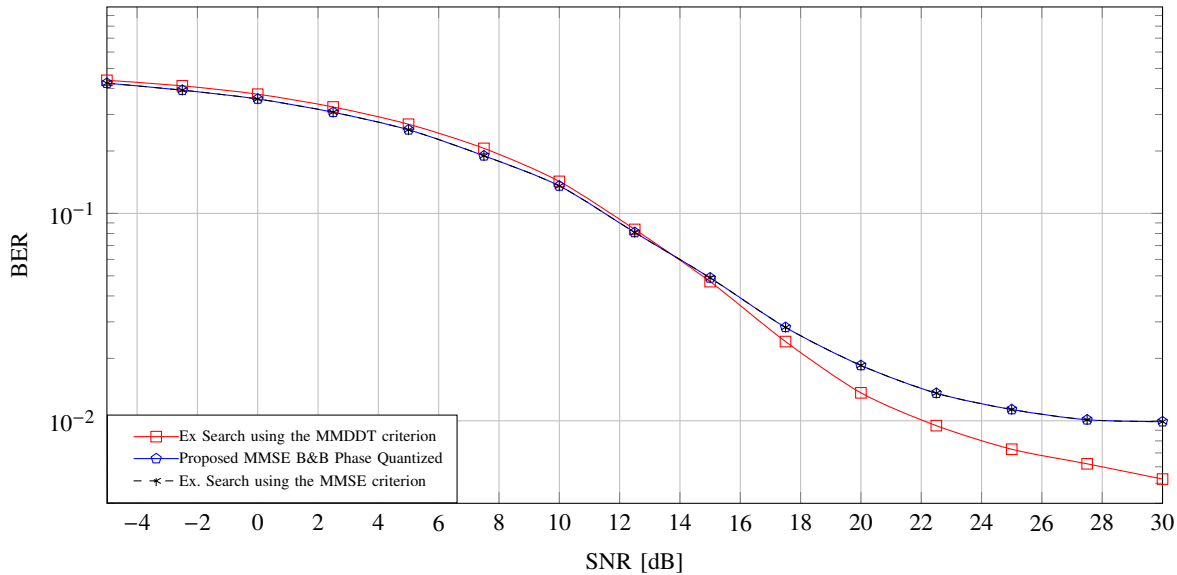
\begin{figure}[ht]
\begin{center}
%
%
%
\usetikzlibrary{positioning,calc}

\definecolor{mycolor1}{rgb}{0.00000,1.00000,1.00000}%
\definecolor{mycolor2}{rgb}{1.00000,0.00000,1.00000}%

\pgfplotsset{every axis label/.append style={font=\footnotesize},
every tick label/.append style={font=\footnotesize}
}

\begin{tikzpicture}[font=\footnotesize] 

\begin{axis}[%
name=ber,
ymode=log,
width  = 0.85\columnwidth,
height = 0.4\columnwidth,
scale only axis,
xmin  = -5,
xmax  = 30,
xlabel= {SNR  [dB]},
xmajorgrids,
ymin=0.004,
ymax=0.99,
ylabel={BER},
ymajorgrids,
legend entries={
                },
legend style={at={(0,0.33)},anchor=south west,draw=black,fill=white,legend cell align=left,font=\tiny}
]

\addlegendimage{solid,no marks,color=black,fill=gray!20,mark=square}


\addplot+[smooth,color=red,solid, every mark/.append style={solid, fill=red!20},mark=square,
y filter/.code={\pgfmathparse{\pgfmathresult-0}\pgfmathresult}]
  table[row sep=crcr]{%
-5	0.438376042	\\
-2.5	0.41231875	\\
0	0.375833333	\\
2.5	0.326036458	\\
5	0.269185417	\\
7.5	0.205861458	\\
10	0.142686458	\\
12.5	0.083708333	\\
15	0.046886458	\\
17.5	0.02410625	\\
20	0.013642708	\\
22.5	0.009475	\\
25	0.007291667	\\
27.5	0.006202083	\\
30	0.005235417	\\
};

\addplot+[smooth,color=blue,solid, every mark/.append style={solid, fill=blue!20},mark=pentagon,
y filter/.code={\pgfmathparse{\pgfmathresult-0}\pgfmathresult}]
  table[row sep=crcr]{%
-5	0.424633333	\\
-2.5	0.393072917	\\
0	0.356351042	\\
2.5	0.307458333	\\
5	0.253032292	\\
7.5	0.189657292	\\
10	0.135705208	\\
12.5	0.080997917	\\
15	0.04889375	\\
17.5	0.028223958	\\
20	0.018495833	\\
22.5	0.013586458	\\
25	0.011336458	\\
27.5	0.010114583	\\
30	0.009902083	\\
};

\addplot+[smooth,color=black,dashed, every mark/.append style={solid, fill=blue!20},mark=asterisk,
y filter/.code={\pgfmathparse{\pgfmathresult-0}\pgfmathresult}]
  table[row sep=crcr]{%
-5	0.424633333	\\
-2.5	0.393072917	\\
0	0.356351042	\\
2.5	0.307458333	\\
5	0.253032292	\\
7.5	0.189657292	\\
10	0.135705208	\\
12.5	0.080997917	\\
15	0.04889375	\\
17.5	0.028223958	\\
20	0.018495833	\\
22.5	0.013586458	\\
25	0.011336458	\\
27.5	0.010114583	\\
30	0.009902083	\\
};

\addplot[smooth,color=red,solid,mark=square,
y filter/.code={\pgfmathparse{\pgfmathresult-0}\pgfmathresult}]
  table[row sep=crcr]{%
	1 2\\
};\label{plot1}

\addplot[smooth,color=blue,solid, mark=pentagon,
y filter/.code={\pgfmathparse{\pgfmathresult-0}\pgfmathresult}]
  table[row sep=crcr]{%
	1 2\\
};\label{plot2}

\addplot[smooth,color=black,dashed, mark=asterisk,
y filter/.code={\pgfmathparse{\pgfmathresult-0}\pgfmathresult}]
  table[row sep=crcr]{%
	1 2\\
};\label{bla}

\node [draw,fill=white,font=\tiny,anchor= south  west] at (axis cs: -5,4*10^-3) {
\setlength{\tabcolsep}{0.5mm}
\renewcommand{\arraystretch}{.8}
\begin{tabular}{l}

\ref{plot1}{Ex Search using the MMDDT criterion}\\
\ref{plot2}{Proposed MMSE B\&B Phase Quantized}\\
\ref{bla}{Ex. Search using the MMSE criterion}\\
\end{tabular}
};

\end{axis}

\end{tikzpicture}%
\caption{Uncoded BER versus $\mathrm{SNR}$ for $K=2$, $M=4$, $\alpha_s=8$ and $\alpha_x=8$} 
\label{fig:Uncoded_BB_precoders}       
\end{center}
\end{figure}
{In this subsection we evaluate the performance of the proposed algorithms with phase quantizers as hard detectors. The analyzed case considers, both, the data and the transmit vector symbols are 8-PSK, which means $\alpha_s=8$ and $\alpha_x=8$.}
Two different scenarios are considered. First we compare the BER performance of the proposed branch-and-bound approach with the branch-and-bound algorithm developed in \cite{General_MMDDT_BB}, which utilizes the MMDDT criterion. The considered system has $K=2$ users and  $M=4$ BS antennas. The BER performances are illustrated in Fig.~\ref{fig:Uncoded_BB_precoders}. \looseness-1

\begin{figure}[ht]
\begin{center}
%
%
%
\usetikzlibrary{positioning,calc}

\definecolor{mycolor1}{rgb}{0.00000,1.00000,1.00000}%
\definecolor{mycolor2}{rgb}{1.00000,0.00000,1.00000}%

\pgfplotsset{every axis label/.append style={font=\footnotesize},
every tick label/.append style={font=\footnotesize}
}

\begin{tikzpicture}[font=\footnotesize] 

\begin{axis}[%
name=ber,
ymode=log,
width  = 0.85\columnwidth,
height = 0.4\columnwidth,
scale only axis,
xmin  = 0,
xmax  = 17.5,
xlabel= {SNR  [dB]},
xmajorgrids,
ymin=0.001,
ymax=1,
ylabel={BER},
ymajorgrids,
legend entries={
                },
legend style={at={(0,0.33)},anchor=south west,draw=black,fill=white,legend cell align=left,font=\tiny}
]

\addlegendimage{solid,no marks,color=black,fill=gray!20,mark=square}


\definecolor{dark_red}{rgb}{0.5,0,0}


\addplot+[smooth,color=gray,solid, every mark/.append style={solid, fill=gray!20},mark=diamond,
y filter/.code={\pgfmathparse{\pgfmathresult-0}\pgfmathresult}]
table[row sep=crcr]{%
-10	     0.448303918650794      \\
-7.5     0.430127678571429      \\
-5	     0.402889285714286      \\
-2.5     0.368207936507937      \\
0	     0.320359771825397      \\
2.5	     0.264776835317460      \\
5	     0.200812946428571      \\
7.5	     0.136242559523810      \\
10	     0.0804849702380952         \\
12.5     0.0394075396825397         \\
15	     0.0161324156746032          \\
17.5     0.00603335813492063                            \\
};

\addplot+[smooth,color=red,solid, every mark/.append style={solid, fill=red!20},mark=square,
y filter/.code={\pgfmathparse{\pgfmathresult-0}\pgfmathresult}]
  table[row sep=crcr]{%
-10	   0.455183258928571        \\
-7.5   0.439056374007937        \\
-5	   0.414327852182540        \\
-2.5   0.382537872023810        \\
0	   0.335609895833333        \\
2.5	   0.279658531746032        \\
5	   0.211894642857143        \\
7.5	   0.142155952380952        \\
10	   0.0796489335317460         \\
12.5   0.0342139880952381         \\
15	   0.0100672371031746           \\
17.5   0.00174764384920635          \\
};

\addplot+[smooth,color=blue,solid, every mark/.append style={solid, fill=blue!20},mark=pentagon,
y filter/.code={\pgfmathparse{\pgfmathresult-0}\pgfmathresult}]
  table[row sep=crcr]{%
-10	    0.448306696428571 \\
-7.5    0.430122048611111 \\
-5	    0.402873387896825 \\
-2.5    0.368193849206349 \\
0	    0.320308506944444 \\
2.5	    0.264447792658730 \\
5	    0.199569518849206 \\
7.5	    0.133418278769841 \\
10	    0.0748756200396825    \\
12.5    0.0324892609126984    \\
15	    0.00974206349206349       \\
17.5    0.00176086309523810       \\
};

\addplot+[smooth,color=black,solid, every mark/.append style={solid, fill=black!20},mark=triangle,
y filter/.code={\pgfmathparse{\pgfmathresult-0}\pgfmathresult}]
  table[row sep=crcr]{%
-10	        0.452733283730159    \\
-7.5        0.436128025793651    \\
-5	        0.410666046626984    \\
-2.5        0.378349007936508    \\
0	        0.332810391865079    \\
2.5	        0.280864384920635    \\
5	        0.223384077380952    \\
7.5	        0.168775768849206    \\
10	        0.123647916666667    \\
12.5        0.0906358382936508        \\
15	        0.0691327628968254        \\
17.5        0.0559528273809524        \\
};

\addplot+[smooth,color=cyan,solid, every mark/.append style={solid, fill=cyan!50},mark=o,
y filter/.code={\pgfmathparse{\pgfmathresult-0}\pgfmathresult}]
  table[row sep=crcr]{%
-10	    0.454053645833333   \\
-7.5    0.437638244047619   \\
-5	    0.412389186507937   \\
-2.5    0.379904563492064   \\
0	    0.332579637896825   \\
2.5	    0.276621502976191   \\
5	    0.210080778769841   \\
7.5	    0.142910887896825   \\
10	    0.0839497271825397    \\
12.5    0.0411525297619048    \\
15	    0.0169179067460317    \\
17.5    0.00634407242063492     \\
};

\addplot+[smooth,color=green,solid, every mark/.append style={solid, fill=green!50},mark=+,
y filter/.code={\pgfmathparse{\pgfmathresult-0}\pgfmathresult}]
  table[row sep=crcr]{%
-10	   0.453965252976191     \\
-7.5   0.437552554563492     \\
-5	   0.412377678571429     \\
-2.5   0.379996701388889     \\
0	   0.333098387896825     \\
2.5	   0.277980431547619     \\
5	   0.213553472222222     \\
7.5	   0.149091269841270     \\
10	   0.0933363095238095         \\
12.5   0.0520585565476190         \\
15	   0.0268556547619048          \\
17.5   0.0136881696428571          \\
};

 \addplot+[smooth,color=orange,solid, every mark/.append style={solid, fill=blue!20},mark=asterisk,
 y filter/.code={\pgfmathparse{\pgfmathresult-0}\pgfmathresult}]
   table[row sep=crcr]{%
 -10     0.439996453373016      \\
 -7.5    0.418382638888889      \\
 -5      0.386253943452381      \\
 -2.5    0.345395758928571      \\
 0	     0.290565649801587      \\
 2.5     0.228252703373016      \\
 5       0.159203497023810      \\
 7.5     0.0940698908730159     \\
 10	     0.0436576388888889     \\
 12.5    0.0139632192460317     \\
 15	     0.00277929067460317        \\
 17.5    0.000276364087301587       \\          
 };             

\addplot+[smooth,color=dark_red,solid, every mark/.append style={solid, fill=gray!20},mark=star,
y filter/.code={\pgfmathparse{\pgfmathresult-0}\pgfmathresult}]
table[row sep=crcr]{%
-10	     0.448515277777778       \\
-7.5     0.430553273809524       \\
-5	     0.403240575396825       \\
-2.5     0.368773561507937       \\
0	     0.321164409722222       \\
2.5	     0.266704092261905       \\
5	     0.205525074404762       \\
7.5	     0.145332490079365       \\
10	     0.0931661954365079            \\
12.5     0.0534217013888889            \\
15	     0.0283784226190476            \\
17.5     0.0147674355158730                  \\
};

\addplot+[smooth,color=magenta,solid, every mark/.append style={solid, fill=gray!20},mark=x,
y filter/.code={\pgfmathparse{\pgfmathresult-0}\pgfmathresult}]
table[row sep=crcr]{%
-10	   0.439645213293651        \\
-7.5   0.419293948412698        \\
-5	   0.392562053571429        \\
-2.5   0.357742385912698        \\
0	   0.315774727182540        \\
2.5	   0.263837524801587        \\
5	   0.202925124007937        \\
7.5	   0.139501884920635        \\
10	   0.0833429811507937            \\
12.5   0.0416195188492064            \\
15	   0.0175490823412698            \\
17.5   0.00690319940476191                   \\
};

\addplot[smooth,color=orange,solid,mark=asterisk,
y filter/.code={\pgfmathparse{\pgfmathresult-0}\pgfmathresult}]
  table[row sep=crcr]{%
	1 2\\
};\label{P0}

\addplot[smooth,color=red,solid,mark=square,
y filter/.code={\pgfmathparse{\pgfmathresult-0}\pgfmathresult}]
  table[row sep=crcr]{%
	1 2\\
};\label{plot7}

\addplot[smooth,color=blue,solid, mark=pentagon,
y filter/.code={\pgfmathparse{\pgfmathresult-0}\pgfmathresult}]
  table[row sep=crcr]{%
	1 2\\
};\label{P6}
\addplot[smooth,color=green,solid,mark=+
,
y filter/.code={\pgfmathparse{\pgfmathresult-0}\pgfmathresult}]
  table[row sep=crcr]{%
	1 2\\
};\label{plot3}

\addplot[smooth,color=black,cyan,mark=o,
y filter/.code={\pgfmathparse{\pgfmathresult-0}\pgfmathresult}]
  table[row sep=crcr]{%
	1 2\\
};\label{plot4}
\addplot[smooth,color=gray,solid,mark=diamond,
y filter/.code={\pgfmathparse{\pgfmathresult-0}\pgfmathresult}]
  table[row sep=crcr]{%
	1 2\\
};\label{plot6}

\addplot[smooth,color=black,solid,mark=triangle,
y filter/.code={\pgfmathparse{\pgfmathresult-0}\pgfmathresult}]
  table[row sep=crcr]{%
	1 2\\
};\label{plot8}

\addplot[smooth,color=dark_red,solid,mark=star,
y filter/.code={\pgfmathparse{\pgfmathresult-0}\pgfmathresult}]
  table[row sep=crcr]{%
	1 2\\
};\label{plot9}

\addplot[smooth,color=magenta,solid,mark=x,
y filter/.code={\pgfmathparse{\pgfmathresult-0}\pgfmathresult}]
  table[row sep=crcr]{%
	1 2\\
};\label{plot10}

\node [draw,fill=white,font=\tiny,anchor= south  west] at (axis cs: 0,0.001) {
\setlength{\tabcolsep}{0.5mm}
\renewcommand{\arraystretch}{.8}
\begin{tabular}{l}

\ref{plot8}{ZF-P Phase quantized \cite{ZF-Precoding}}\\
\ref{plot3}{CVX-CIO Phase quantized \cite{CVX-CIO}}\\
\ref{plot9}{SQUID Phase quantized (Single-Carrier) \cite{jacobsson2018nonlinear}}\\
\ref{plot10}{C3PO Phase quantized \cite{Struder_c3po}} \\
\ref{plot4}{MSM Phase quantized \cite{MSM_precoder}}\\
\ref{plot6}{Proposed MMSE Mapped Phase quantized}\\
\ref{plot7}{MMDDT B\&B Phase quantized} \cite{Landau2017}\\
\ref{P6}{Proposed MMSE B\&B Phase quantized}\\
\ref{P0}{Linear MMSE Unquantized \cite{M_Joham_ZF}}
\end{tabular}
};

\end{axis}

\end{tikzpicture}%
\captionsetup{justification=centering}
\caption{Uncoded BER versus $\mathrm{SNR}$, $K=3$, $M=12$, $\alpha_s=8$ and $\alpha_x=8$} 
\label{fig:Uncoded_all_precoders}       
\end{center}
\end{figure}
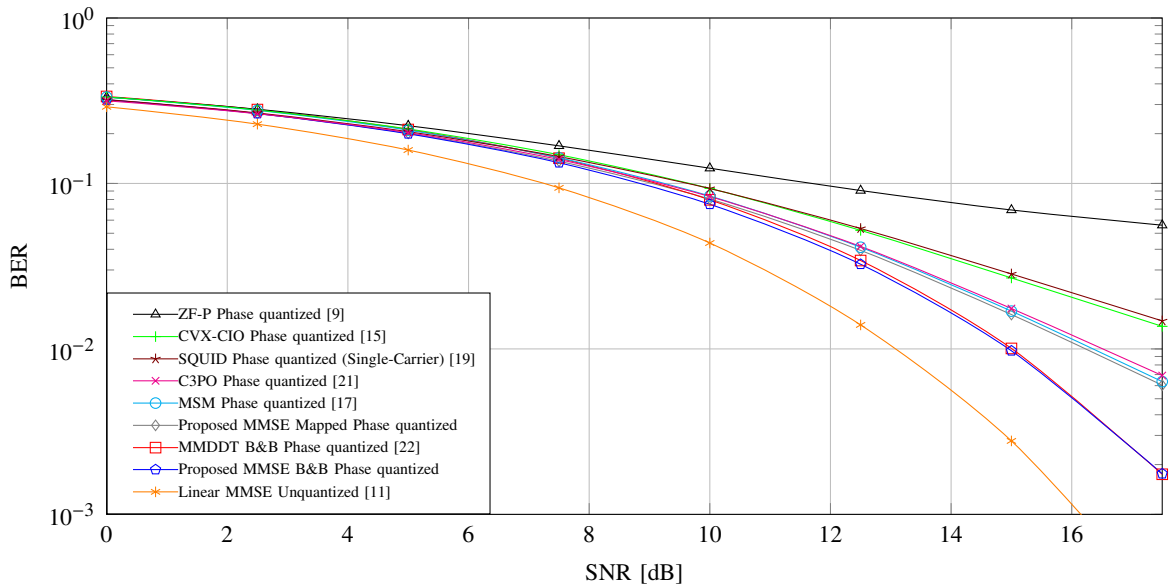

In the second scenario, we consider { $K=3$ users and the number of antennas at the BS $M=12$} and compare both proposed methods with state-of-the-art approaches. The BER performances
are illustrated in Fig.~\ref{fig:Uncoded_all_precoders}.

Fig.~\ref{fig:Uncoded_BB_precoders} confirms the superiority of the MMSE criterion against MMDDT for low SNR. On the other hand, Fig.~\ref{fig:Uncoded_BB_precoders} shows that the MMDDT criterion is favorable for the high SNR regime where it yields a marginally lower uncoded BER.

The results shown in Fig.~\ref{fig:Uncoded_all_precoders} illustrate a significant gain in BER performance when using the optimal MMSE branch-and-bound method in contrast to suboptimal low resolution approaches. Moreover, the proposed branch-and-bound precoder shows only a 2 dB loss in comparison with the full resolution MMSE linear precoding strategy presented in \cite{M_Joham_ZF}. 
Fig.~\ref{fig:Uncoded_all_precoders} also confirms
the suitability of the MMSE criterion for low and medium SNR once the proposed optimal approach outperforms all other low resolution schemes for that SNR regime, in terms of BER. Besides that, the results indicate that the proposed suboptimal approach termed MMSE Mapped surprisingly outperforms other suboptimal state-of-the-art algorithms in terms of BER performance.

\subsubsection{Complexity Analysis}
$\\$
\textcolor{r}{In this subsection we evaluate the average computational complexity of the proposed algorithms and compare them against the complexity of the state-of-the-art approaches considering the same scenario presented in the previous subsection, meaning $K=3$ users $M=12$ BS antennas and 8-PSK data and transmit vector symbols. The complexity is evaluated in terms of the complexity order and also in terms of the execution time.}

\paragraph{\textcolor{r}{Complexity Order Analysis}}
$\\$
The computational complexity of each algorithm is summarized in Table~\ref{tab:table1}, where $B$ denotes the number of evaluated bounds in the corresponding branch-and-bound algorithm and \textcolor{r}{$T$ is the number of iterations of a particular algorithm}. 

\begin{table}[ht]
\centering
{\scriptsize{
  \caption{Computational Complexity of the Algorithms}
  \label{tab:table1}
  \begin{tabular}{ | p {15em} | m {23em}| p{16em}|}
  \hline
    {Algorithm} & {Complexity} &{Optimization Problem Type}\\ \hline \hline
    MSM-Precoder \cite{MSM_precoder}& {$\mathcal{O}((2M+1)^{3.5})$} &Linear Program\\ \hline
    ZF-P \cite{ZF-Precoding}& {$\mathcal{O}(K^{2}M)$} &-\\ \hline
    CVX-CIO \cite{CVX-CIO} & {$\mathcal{O}((2M)^{3.5})$} &Second Order Cone Program\\ \hline
    SQUID-OFDM \cite{jacobsson2018nonlinear}  & {$\mathcal{O}\pc{\frac{10}{3}K^3+6 MK^2+12KM-\frac{4}{3}K+T(8MK+4M)}$} &-\\ \hline
    C3PO \cite{Struder_c3po}& {$\mathcal{O}(4(M^2(K+1)+T M^2+ MK+K^2))$} &Non Convex Quadratic Program\\ \hline
    MMDDT branch-and-bound \cite{General_MMDDT_BB}  & {$\mathcal{O}({B} \hspace{0.1em}(2M+1)^{3.5})$} &Discrete Program\\ \hline
    Proposed MMSE branch-and-bound  & {$\mathcal{O}({B} \hspace{0.1em}(2M+1)^{3.5})$} &Discrete Program\\ \hline
    Proposed MMSE Mapped  & {$\mathcal{O}((2M+1)^{3.5})$} &Quadratic Program\\ \hline
  \end{tabular}
}}
\end{table}
\begin{figure}[ht]
\begin{center}
%
%
%
\usetikzlibrary{positioning,calc}

\definecolor{mycolor1}{rgb}{0.00000,1.00000,1.00000}%
\definecolor{mycolor2}{rgb}{1.00000,0.00000,1.00000}%

\pgfplotsset{every axis label/.append style={font=\footnotesize},
every tick label/.append style={font=\footnotesize}
}

\begin{tikzpicture}[font=\footnotesize] 

\begin{axis}[%
name=A ,
ymode=log,
width  = 0.3\columnwidth,
height = 0.25\columnwidth,
scale only axis,
xmin  = -10,
xmax  = 17.5,
xlabel= {$\text{SNR}$},
xmajorgrids,
ymin=5,
ymax=8^13,
ylabel={$B$ },
ymajorgrids,
legend entries={Ex. Search,
				MMSE-B\&B,
				MMDDT-B\&B    \cite{Landau2017}
                },
legend style={at={(0,0.75)},anchor=north west,draw=black,fill=white,legend cell align=left,font=\tiny}
]

\addlegendimage{color=black,dashed,fill=gray!20,mark=asterisk}
\addlegendimage{color=blue,fill=gray!20,mark=pentagon}
\addlegendimage{color=red,fill=gray!20,mark=square}


\addplot+[smooth,color=blue,solid, every mark/.append style={solid, fill=cyan!20},mark=pentagon,
y filter/.code={\pgfmathparse{\pgfmathresult-0}\pgfmathresult}]
  table[row sep=crcr]{%
-10	    19.4487500000000    \\
-7.5    30.3187500000000    \\
-5	    45.2312500000000    \\
-2.5    62.8275000000000    \\
0	    84.2087500000000    \\
2.5	    112.731250000000    \\
5	    156.571250000000    \\
7.5	    236.181250000000    \\
10	    373.013750000000    \\
12.5    618.670000000000    \\
15	    1100.88375000000    \\
17.5    2079.12625000000    \\
};

\addplot+[smooth,color=red,solid, every mark/.append style={solid, fill=blue!20},mark=square,
y filter/.code={\pgfmathparse{\pgfmathresult-0}\pgfmathresult}]
  table[row sep=crcr]{%
-10	    9779.97375000000 \\
-7.5    9779.97375000000 \\
-5	    9779.97375000000 \\
-2.5    9779.97375000000 \\
0	    9779.97375000000 \\
2.5	    9779.97375000000 \\
5	    9779.97375000000 \\
7.5	    9779.97375000000 \\
10	    9779.97375000000 \\
12.5    9779.97375000000 \\
15	    9779.97375000000 \\
17.5    9779.97375000000 \\
};

%

\addplot+[smooth,color=black, dashed, every mark/.append style={solid, fill=cyan!20},mark=asterisk,
y filter/.code={\pgfmathparse{\pgfmathresult-0}\pgfmathresult}]
  table[row sep=crcr]{%
-10	        68719476736.0000        \\
-7.5        68719476736.0000        \\
-5	        68719476736.0000        \\
-2.5        68719476736.0000        \\
0	        68719476736.0000        \\
2.5	        68719476736.0000        \\
5	        68719476736.0000        \\
7.5	        68719476736.0000        \\
10	        68719476736.0000        \\
12.5        68719476736.0000        \\
15	        68719476736.0000        \\
17.5        68719476736.0000        \\
20	        68719476736.0000        \\
22.5        68719476736.0000        \\
25	        68719476736.0000        \\
27.5        68719476736.0000        \\
30	        68719476736.0000        \\
};

%
%
%
%
%
%
%
%

\end{axis}

\begin{axis}[%
name=SumRate,
at={($(A.east)+(40,0em)$)},
                anchor= west,
name=B ,
ymode=log,
width  = 0.3\columnwidth,
height = 0.25\columnwidth,
scale only axis,
xmin  = 3,
xmax  = 18,
xlabel= {$M$},
xmajorgrids,
ymin=0,
ymax=10000000,
ylabel={$B$ },
ymajorgrids,
]

\addlegendimage{color=black,dashed,fill=gray!20,mark=asterisk}
\addlegendimage{color=blue,fill=gray!20,mark=pentagon}
\addlegendimage{color=red,fill=gray!20,mark=square}



\addplot+[smooth,color=black,dashed, every mark/.append style={solid, fill=gray!20},mark=asterisk,
y filter/.code={\pgfmathparse{\pgfmathresult-0}\pgfmathresult}]
  table[row sep=crcr]{%
3	512                   \\ 
4	4096                  \\     
5	32768                 \\     
6	262144                \\     
7	2097152               \\     
8	16777216              \\         
9	134217728             \\         
10	1073741824.00000      \\                 
11	8589934592.00000      \\                 
12	68719476736.0000      \\                 
13	549755813888.000      \\                 
14	4398046511104.00      \\                 
15	35184372088832.0      \\                 
16	281474976710656       \\             
17	2.25179981368525e+15  \\                     
18	1.80143985094820e+16  \\                     
19	1.44115188075856e+17  \\                     
20	1.15292150460685e+18  \\                     
};

\addplot+[smooth,color=red,solid, every mark/.append style={solid, fill=blue!20},mark=square,
y filter/.code={\pgfmathparse{\pgfmathresult-0}\pgfmathresult}]
  table[row sep=crcr]{%
3	5.01500000000000	            \\
4	13.9353125000000	            \\
5	28.2096875000000	            \\
6	58.1696875000000	            \\
7	113.123281250000	            \\
8	186.555937500000	            \\
9	300.815156250000	            \\
10	497.200781250000	            \\
11	853.523906250000	            \\
12	1222.49671875000	            \\
13	1804.15765625000                \\
14	2432.70984375000                \\
15	3663.27234375000                \\
16	5017.04062500000	        \\
17	7234.65156250000    \\
18	9580.55000000000    \\
19	2618.295	\\ 
20	3225.04	\\
};

\addplot+[smooth,color=blue,solid, every mark/.append style={solid, fill=cyan!20},mark=pentagon,
y filter/.code={\pgfmathparse{\pgfmathresult-0}\pgfmathresult}]
  table[row sep=crcr]{%
3	1.51032812500000    \\
4	2.40187500000000    \\
5	3.46804687500000    \\
6	4.57025000000000    \\
7	5.85831250000000    \\
8	7.28776562500000    \\
9	8.81282812500000    \\
10	10.4024062500000    \\
11	12.2242812500000    \\
12	14.1094375000000    \\
13	16.2112656250000    \\
14	18.1065468750000    \\
15	20.5295000000000    \\
16	22.9001093750000    \\
17	25.1570156250000    	\\
18	28.2286562500000    	\\
19	30.5873593750000    	\\
20	33.5222656250000    	\\
};

\addplot[smooth,color=red,dashed,
y filter/.code={\pgfmathparse{\pgfmathresult-0}\pgfmathresult}]
  table[row sep=crcr]{%
	-20 2\\
};\label{P99}
\addplot[smooth,color=black,solid,
y filter/.code={\pgfmathparse{\pgfmathresult-0}\pgfmathresult}]
  table[row sep=crcr]{%
	-20 2\\
};\label{P88}


\end{axis}

\end{tikzpicture}%
\caption{Average number of evaluated bounds $\times$ SNR, $K=3$, $M=12 \ , \alpha_s=\alpha_x=8$ (left).
Average number of evaluated bounds $\times$  Number of transmit antennas, $K=3$, $\text{SNR}=3 \text{dB}, \ \alpha_s=\alpha_x=8$  (right).} 
\label{fig:Complexity}       
\vspace{-2em}
\end{center}
\end{figure}
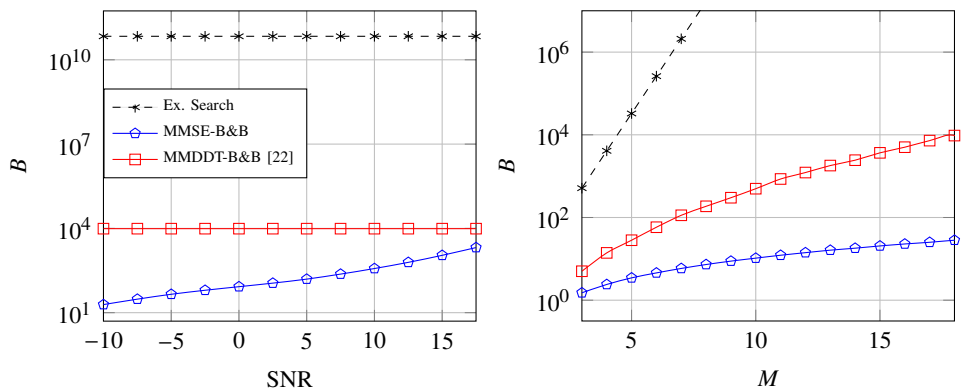 
The results shown in Table~\ref{tab:table1} were obtained considering that by using the interior points method (IPM) the optimization problems of the state-of-the-art algorithms can be solved with complexity in the order of $\mathcal{O}(n^{3.5})$, with $n\leq(2M+1)$, cf.\ \cite{Boyd_2004}.

Moreover, Fig.~\ref{fig:Complexity} shows the number of evaluated bounds of the proposed branch-and-bound method compared with the ones from the study from \cite{General_MMDDT_BB} and exhaustive search for the considered system.
In Fig.~\ref{fig:Complexity} it can be seen that the number of evaluated bounds of the proposed branch-and-bound method is significantly smaller than the one from \cite{General_MMDDT_BB} for low SNR, which underlines the superiority of our proposed method to the existing ones for low SNR.
Furthermore, based on Fig.~\ref{fig:Complexity},  the average number of subproblems solved is always significantly smaller than the total number of candidates to be evaluated in the exhaustive search. \textcolor{r}{Taking into account that each candidate evaluation in the exhaustive search corresponds to a complexity in the order of $\mathcal{O}(KM)$ justifies the use of the proposed method when the optimal precoding vector is desired.}

\begin{figure}[ht]
\begin{center}
%
%
%
\usetikzlibrary{positioning,calc}

\definecolor{mycolor1}{rgb}{0.00000,1.00000,1.00000}%
\definecolor{mycolor2}{rgb}{1.00000,0.00000,1.00000}%

\pgfplotsset{every axis label/.append style={font=\footnotesize},
every tick label/.append style={font=\footnotesize}
}

\begin{tikzpicture}[font=\footnotesize] 

\begin{axis}[%
name=ber,
ymode=log,
width  = 0.85\columnwidth,
height = 0.4\columnwidth,
scale only axis,
xmin  = -10,
xmax  = 17.5,
xlabel= {SNR  [dB]},
xmajorgrids,
ymin=10^-2,
ymax=5*10^6,
ylabel={Execution Time [ms]},
ymajorgrids,
legend entries={,Proposed MMSE Mapped Phase quantized,
MMDDT B\&B Phase quantized \cite{General_MMDDT_BB},
Proposed MMSE B\&B Phase quantized,
ZF-P Phase quantized \cite{ZF-Precoding},
MSM Phase quantized \cite{MSM_precoder},
CVX-CIO Phase quantized \cite{CVX-CIO},
Linear MMSE Unquantized \cite{M_Joham_ZF},
SQUID Phase quantized (Single-Carrier) \cite{jacobsson2018nonlinear},
C3PO Phase quantized \cite{Struder_c3po}, 
},
legend columns=3,
legend style={at={(0,1)},anchor=north west,draw=black,fill=white,legend cell align=left,font=\tiny,}
]

\addlegendimage{solid,no marks,color=black,fill=gray!20,mark=square}


\definecolor{dark_red}{rgb}{0.5,0,0}


\addplot+[smooth,color=gray,solid, every mark/.append style={solid, fill=gray!20},mark=diamond,
y filter/.code={\pgfmathparse{\pgfmathresult-0}\pgfmathresult}]
table[row sep=crcr]{%
-10	         3.0429    \\
-7.5         2.8491    \\
-5	         2.8636    \\
-2.5         2.9788    \\
0	         3.0238    \\
2.5	         2.9359    \\
5	         3.0095    \\
7.5	         2.9992    \\
10	         3.0722        \\
12.5         3.1920        \\
15	         3.2503         \\
17.5         3.3616                            \\
};

\addplot+[smooth,color=red,solid, every mark/.append style={solid, fill=red!20},mark=square,
y filter/.code={\pgfmathparse{\pgfmathresult-0}\pgfmathresult}]
  table[row sep=crcr]{%
-10	  27799.2334200625      \\
-7.5  27799.2334200625      \\
-5	  27799.2334200625      \\
-2.5  27799.2334200625      \\
0	  27799.2334200625      \\
2.5	  27799.2334200625      \\
5	  27799.2334200625      \\
7.5	  27799.2334200625      \\
10	  27799.2334200625        \\
12.5  27799.2334200625        \\
15	  27799.2334200625          \\
17.5  27799.2334200625          \\
};

\addplot+[smooth,color=blue,solid, every mark/.append style={solid, fill=blue!20},mark=pentagon,
y filter/.code={\pgfmathparse{\pgfmathresult-0}\pgfmathresult}]
  table[row sep=crcr]{%
-10	        43.5765518593750      \\
-7.5        64.4474717968750      \\
-5	        94.6837696562500      \\
-2.5        129.855075078125     \\
0	        171.503520296875     \\
2.5	        226.111630500000     \\
5	        310.022713703125     \\
7.5	        461.770596218750     \\
10	        723.213501484375     \\
12.5        1192.58542690625    \\
15	        2109.22202996875    \\
17.5        3953.79796479688    \\
};

\addplot+[smooth,color=black,solid, every mark/.append style={solid, fill=black!20},mark=triangle,
y filter/.code={\pgfmathparse{\pgfmathresult-0}\pgfmathresult}]
  table[row sep=crcr]{%
-10	     0.0449   \\
-7.5     0.0449   \\
-5	     0.0449   \\
-2.5     0.0449   \\
0	     0.0449   \\
2.5	     0.0449   \\
5	     0.0449   \\
7.5	     0.0449   \\
10	     0.0449   \\
12.5     0.0449        \\
15	     0.0449        \\
17.5     0.0449        \\
};

\addplot+[smooth,color=cyan,solid, every mark/.append style={solid, fill=cyan!50},mark=o,
y filter/.code={\pgfmathparse{\pgfmathresult-0}\pgfmathresult}]
  table[row sep=crcr]{%
-10	        9.3429   \\
-7.5        9.3429   \\
-5	        9.3429   \\
-2.5        9.3429   \\
0	        9.3429   \\
2.5	        9.3429   \\
5	        9.3429   \\
7.5	        9.3429   \\
10	        9.3429     \\
12.5        9.3429     \\
15	        9.3429     \\
17.5        9.3429       \\
};

\addplot+[smooth,color=green,solid, every mark/.append style={solid, fill=green!50},mark=+,
y filter/.code={\pgfmathparse{\pgfmathresult-0}\pgfmathresult}]
  table[row sep=crcr]{%
-10	  1046.71432954688    \\
-7.5  1046.71432954688    \\
-5	  1046.71432954688    \\
-2.5  1046.71432954688    \\
0	  1046.71432954688    \\
2.5	  1046.71432954688    \\
5	  1046.71432954688    \\
7.5	  1046.71432954688    \\
10	  1046.71432954688         \\
12.5  1046.71432954688         \\
15	  1046.71432954688          \\
17.5  1046.71432954688          \\
};

 \addplot+[smooth,color=orange,solid, every mark/.append style={solid, fill=blue!20},mark=asterisk,
 y filter/.code={\pgfmathparse{\pgfmathresult-0}\pgfmathresult}]
   table[row sep=crcr]{%
 -10     0.371668359375000   \\
 -7.5    0.103526093750000   \\
 -5      0.107977156250000   \\
 -2.5    0.127287000000000   \\
 0	     0.137699390625000   \\
 2.5     0.137032140625000   \\
 5       0.130785687500000   \\
 7.5     0.132598671875000   \\
 10	     0.139727156250000   \\
 12.5    0.121737671875000   \\
 15	     0.152972296875000       \\
 17.5    0.139645609375000       \\          
 };             

\addplot+[smooth,color=dark_red,solid, every mark/.append style={solid, fill=gray!20},mark=star,
y filter/.code={\pgfmathparse{\pgfmathresult-0}\pgfmathresult}]
table[row sep=crcr]{%
-10	     1.66037735937500     \\
-7.5     0.794222515625000     \\
-5	     0.776955890625000     \\
-2.5     0.800363734375000     \\
0	     0.789401343750000     \\
2.5	     0.790339875000000     \\
5	     0.693195656250000     \\
7.5	     0.641656906250000     \\
10	     0.640170500000000           \\
12.5     0.606028562500000           \\
15	     0.610578640625000           \\
17.5     0.595467750000000                 \\
};

\addplot+[smooth,color=magenta,solid, every mark/.append style={solid, fill=gray!20},mark=x,
y filter/.code={\pgfmathparse{\pgfmathresult-0}\pgfmathresult}]
table[row sep=crcr]{%
-10	   1.93209821875000      \\
-7.5   1.93209821875000      \\
-5	   1.93209821875000      \\
-2.5   1.93209821875000      \\
0	   1.93209821875000      \\
2.5	   1.93209821875000      \\
5	   1.93209821875000      \\
7.5	   1.93209821875000      \\
10	   1.93209821875000           \\
12.5   1.93209821875000           \\
15	   1.93209821875000           \\
17.5   1.93209821875000                   \\
};

\end{axis}

\end{tikzpicture}%
\caption{Execution Time (ms) versus $\mathrm{SNR}$, $K=3$, $M=12$, $\alpha_s=8$ and $\alpha_x=8$ }
\label{fig:Runtime}       
\end{center}
\end{figure}
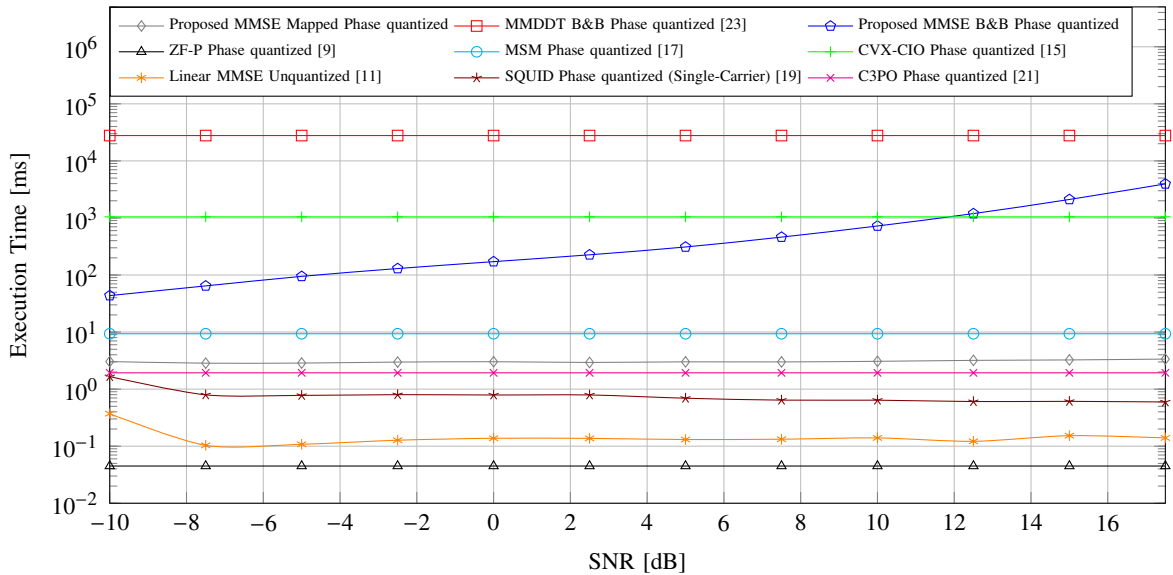

\paragraph{\textcolor{r}{Execution Time Analysis}}
$\\$
\textcolor{r}{In this subsection we evaluate the proposed methods against the state-of-the-art approaches in terms of the execution time. For solving the optimization problems for the methods in \cite{MSM_precoder}, \cite{General_MMDDT_BB}, the proposed MMSE Mapped and the proposed MMSE branch-and-bound the optimization toolbox from Matlab was utilized. The optimization problem required for the CVX-CIO approach \cite{CVX-CIO} was solved via CVX \cite{Boyd_2004}, as suggested by the authors. The results are shown in Fig.~\ref{fig:Runtime}.}
\textcolor{r}{As can be seen in Fig.\ref{fig:Runtime} the proposed MMSE Mapped method has similar execution time as most state-of-the-art suboptimal precoding techniques. Also the proposed MMSE branch-and-bound approach outperforms its the MMDDT version for all evaluated values of SNR and for low and intermediate SNR regime also outperforms the CVX-CIO method.}

\subsection{Coded Transmission}

In this subsection, the proposed soft detection schemes are evaluated considering as the proposed MMSE branch-and-bound approach the precoding technique.
The proposed soft detection schemes are evaluated against the method for extrinsic information computation in AWGN channels described by
\begin{align}
\label{eq:extrinsic_awgn}
   L_e\pc{c_{k,i}}=
    \text{ln}\pc{
    \frac{\displaystyle \sum_{s\in S_{+1}}  \text{e}^{-\frac{{\PM{z_k[t]-s}}^2}{{\sigma}_{w}^2}} \ {P}\pc{s|{r}_{k,t,\upsilon} = +1}
}
    {\displaystyle\sum_{s\in S_{-1}}  \text{e}^{-\frac{{\PM{z_k[t]-s}}^2}{{\sigma}_{w}^2}}\ {P}\pc{s|{r}_{k,t,\upsilon} = -1}}
    },
\end{align}
\textcolor{r}{similar as considered in \cite{jacobsson2018nonlinear} in the context of convolutional codes with a BCJR decoder.}
The analysis is made in conjunction with the IDD scheme presented in subsection \ref{subsec:dpa_idd_algorithm}.

The shown results were computed using an LDPC block code with a block size of $\frac{N_b}{R}=486$ bits and code rate $R=1/2$. The LLRs are processed by sum-product algorithm (SPA) decoders \cite{SPA-Decoder}. The examined system has $K=2$ users and $B=6$ BS antennas where the data symbols are considered as 8-PSK and the precoded symbols are considered as QPSK, meaning $\alpha_s=8$ and $\alpha_x=4$. 
The examined approaches are as follows
\begin{itemize}
    \item[1.] Uncoded transmission
    \item[2.] Coded transmission using the DPA soft detector \eqref{eq:optimal_llr} for the computation of $\boldsymbol{L}_e$;\looseness-1
    \item[3.] Coded transmission using the GDPA soft detector \eqref{eq:LLR_GDPA} for the computation of $\boldsymbol{L}_e$;\looseness-1
    \item[4.] Coded transmission using DPA-LM soft detector \eqref{eq:linear_model_LLR} for the computation of $\boldsymbol{L}_e$;\looseness-1
    \item[5.] Coded transmission using AWGN method \cite{jacobsson2018nonlinear} for the computation of $\boldsymbol{L}_e$.\looseness-1
\end{itemize}

\begin{figure}[ht]
\input{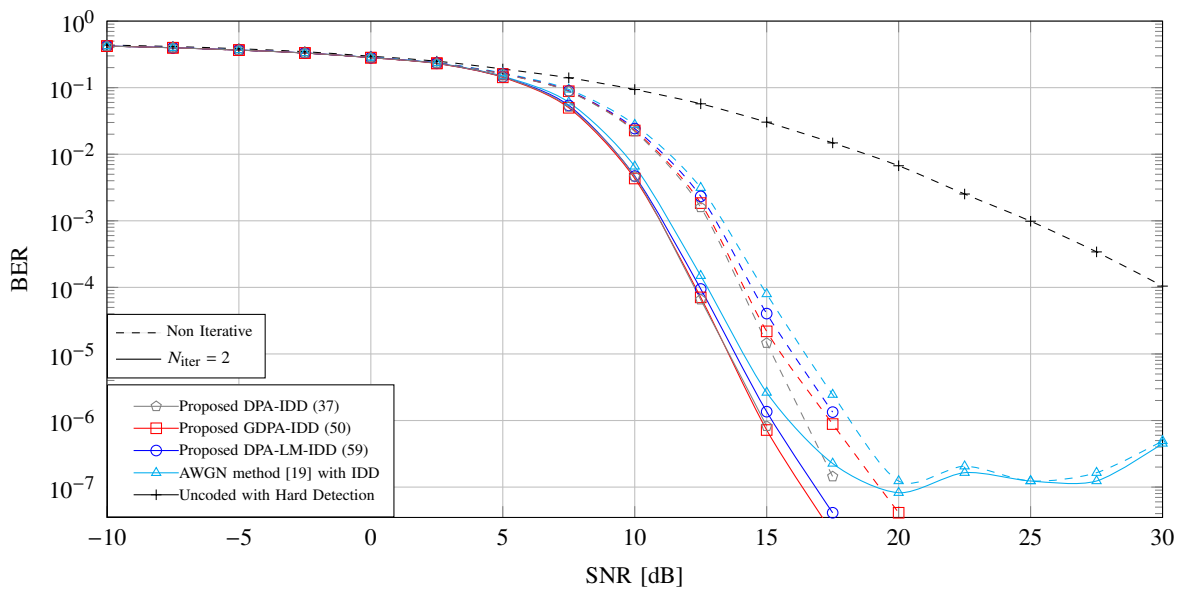}
\caption{Coded BER versus $\mathrm{SNR},\ K=2,\ M=6, \ \alpha_s=8, \ \alpha_x=4$} 
\label{fig:BER_K2_M10}       
\end{figure}

{As can be seen in Fig.~\ref{fig:BER_K2_M10}, all proposed methods provide similar performance for low-SNR. As expected, for the high-SNR regime the proposed DPA-IDD method, that relies on the true PDF of the received signal, yields a lower BER as compared with the proposed suboptimal methods.
Furthermore, considering the marginal performance loss referring to the proposed DPA-IDD method, shown in Fig.~\ref{fig:BER_K2_M10}, reasonable complexity performance trade-offs can be achieved via using the proposed suboptimal methods GDPA-IDD and DPA-LM-IDD.}

{The BER performance associated with the system that uses the common AWGN soft detector is similar to the proposed methods for low-SNR. However, in the medium and high-SNR regime, the distortion brought by the discrete precoding becomes relevant, and, since this is not considered in the common AWGN receive processing it results in an error floor in the BER, as shown in Fig.~\ref{fig:BER_K2_M10}.} \looseness-1

{Finally, Fig.~\ref{fig:BER_K2_M10} shows an improvement in performance when using the iterative method. With a relatively small number of iterations there is a gain of approximately $1.5 \ \text{dB}$ when compared with the non iterative approach.}
\section{Conclusions}
\label{sec:conclusion}

We propose an optimal MMSE precoding technique using quantized signals with constant envelope and PSK modulation. Unlike the existing MMSE design for 1-bit resolution \cite{Jacobsson2018}, the proposed approach employs uniform phase quantization and the bounding step in the branch-and-bound method is different in terms of considering the most restrictive relaxation of the nonconvex problem, which is then utilized for a suboptimal design also. 

The proposed optimal approach outperforms the existing methods in terms of BER for the low and medium SNR regime, with less computational complexity when compared with the branch-and-bound method that utilizes the established MMDDT criterion. 

\textcolor{r}{Furthermore, the proposed suboptimal approach yields similar BER performance when compared with the MSM and C3PO approaches.
At the same time the proposed suboptimal approach provides more flexibility than the C3PO algorithm (3-bit) and it yields a benefit in terms of runtime when compared with the MSM algorithm. With this, it can be understood as a practical solution for systems that utilize discrete precoding.}

Moreover, unlike prior works, the present study proposes three sophisticated soft detectors which allow channel coding schemes to be used in conjunction with low-resolution precoding. Results based on LDPC block coding indicate superior performance of the proposed receive processing as compared to the conventional system design in terms of BER.

\appendix

\subsection{The Conventional MMSE Cost Function With the Scaling Factor}
\label{app:Hessian_MMSE}
The corresponding real valued function of the equivalent MMSE cost function including the scaling factor reads as
\begin{align}
J( \boldsymbol{x}_{\text{r}}, f)= &f^2 \boldsymbol{x}_{\text{r}}^T \boldsymbol{H}_{\text{r}}^T \boldsymbol{H}_{\text{r}}\boldsymbol{x}_{\text{r}}  -2 f    \boldsymbol{x}_{\text{r}}^T \boldsymbol{H}_{\text{r}}^T \boldsymbol{s}_{\text{r}} +
f^2 \mathrm{E} \{ \boldsymbol{w}_{\text{r}}^T\boldsymbol{w}_{\text{r}}  \} \text{.}
\end{align}
The Hessian is constructed based on the partial derivatives given by 
\begin{align}
\boldsymbol{\Gamma}  & =
\frac{ \partial^2 J(\boldsymbol{x}_{\text{r}}, f)}{ \partial  \boldsymbol{x}_{\text{r}}  \partial  \boldsymbol{x}_{\text{r}}^T  }
  = 
2 f^2 
   \boldsymbol{H}_{\text{r}}^T
   \boldsymbol{H}_{\text{r}}  \text{,}   \\
\epsilon & =\frac{ \partial^2 J(\boldsymbol{x}_{\text{r}}, f)}{ \partial  f^2 }
 = 
  2 \lVert  
\boldsymbol{H}_{\text{r}} \boldsymbol{x}_{\text{r}}
    \rVert_2^2  
 +
 2 
   \mathrm{E} \{  \boldsymbol{w}_{\text{r}}^T
   \boldsymbol{w}_{\text{r}}
  \}    \geq 0   \text{,} \notag \\
\boldsymbol{\eta}  & =
\frac{ \partial^2 J(\boldsymbol{x}_{\text{r}}, f)}{ \partial  \boldsymbol{x}_{\text{r}}  \partial  f }
  =  
   4 f 
   \boldsymbol{H}_{\text{r}}^T
   \boldsymbol{H}_{\text{r}}
   \boldsymbol{x}_{\text{r}} 
      -
   2\boldsymbol{H}_{\text{r}}^T  \boldsymbol{s}_{\text{r}}  \notag  \text{.}
\end{align}
Positive semi-definiteness of the Hessian is established when the following inequality holds
\begin{align}
\boldsymbol{v}^T \boldsymbol{\Gamma}
\boldsymbol{v}    
+  \nu^2 \epsilon +
2 \nu \boldsymbol{\eta}^T \boldsymbol{v}
\geq 0 \ \ \ \ \text{for all  } \nu, \boldsymbol{v}    \text{.}
\label{eq:hessian_scalar_common}
\end{align}
Assuming that  $\boldsymbol{\Gamma}\succ \boldsymbol{0}$, the minimum value of the LHS of \eqref{eq:hessian_scalar_common} is given by
$\nu^2 \epsilon - \nu^2 \boldsymbol{\eta}^T \boldsymbol{\Gamma}^{-1}   \boldsymbol{\eta} $. For positive semi-definiteness of the Hessian, this minimum must be greater or equal to zero which yields the condition
\begin{align}
 \epsilon  \geq  \boldsymbol{\eta}^T \boldsymbol{\Gamma}^{-1}   \boldsymbol{\eta} \text{.} \label{eq:epsilon_ieq}
\end{align}
Inserting the partial derivatives in \eqref{eq:epsilon_ieq} gives
\begin{align}
   &2 \lVert  
\boldsymbol{H}_{\text{r}} \boldsymbol{x}_{\text{r}}
    \rVert_2^2  
 +
 2 
   \mathrm{E} \{  \boldsymbol{w}_{\text{r}}^T
   \boldsymbol{w}_{\text{r}}
  \}   \geq  \frac{1}{2 f^2}  \boldsymbol{\eta}^T   (\boldsymbol{H}_{\text{r}}^T \boldsymbol{H}_{\text{r}} ) ^{-1}   \boldsymbol{\eta}   \text{,} \\
   &2 \lVert  
\boldsymbol{H}_{\text{r}} \boldsymbol{x}_{\text{r}}
    \rVert_2^2  
 +
 2 
   \mathrm{E} \{  \boldsymbol{w}_{\text{r}}^T
   \boldsymbol{w}_{\text{r}}
  \}   \geq    8  \boldsymbol{x}_{\text{r}}^T \boldsymbol{H}_{\text{r}}^T \boldsymbol{H}_{\text{r}} \boldsymbol{x}_{\text{r}}    - \frac{8}{f}  \boldsymbol{x}_{\text{r}}^T \boldsymbol{H}^T_{\text{r}}  \boldsymbol{s}_{\text{r}}  +  \boldsymbol{s}_{\text{r}}^T\boldsymbol{H}_{\text{r}}(\boldsymbol{H}_{\text{r}}^T \boldsymbol{H}_{\text{r}} ) ^{-1} \boldsymbol{H}_{\text{r}}^T   \boldsymbol{s}_{\text{r}} \text{,}
\end{align}
which can be rearranged as
\begin{align}
\frac{8}{f}  \boldsymbol{x}_{\text{r}}^T \boldsymbol{H}^T_{\text{r}}  \boldsymbol{s}_{\text{r}}
        \notag  \geq  
    &6  \boldsymbol{x}_{\text{r}}^T \boldsymbol{H}_{\text{r}}^T \boldsymbol{H}_{\text{r}} \boldsymbol{x}_{\text{r}}  +
    2 \boldsymbol{s}_{\text{r}}^T\boldsymbol{H}_{\text{r}}(\boldsymbol{H}_{\text{r}}^T \boldsymbol{H}_{\text{r}} ) ^{-1} \boldsymbol{H}_{\text{r}}^T   \boldsymbol{s}_{\text{r}}
    -
    2 
   \mathrm{E} \{  \boldsymbol{w}_{\text{r}}^T
   \boldsymbol{w}_{\text{r}}  \}
    \text{.} \label{eq:convex_MMSE_not}
\end{align}
We conclude that the MMSE cost function is in general not jointly convex in $\boldsymbol{x}_{\text{r}}$ and $f$.

\subsection{The Partial MMSE Cost Function}
\label{app:Partial_MMSE}
The MMSE cost function for the problem formulation with the $f^{\prime}$ scaled polyhedron and partially fixed precoding vector is given by
\begin{align}
J\pc{\boldsymbol{x}_{\text{r}, f}^{\prime}, f^{\prime}}=&
\PM{ \PM{  \boldsymbol{H}_{\text{r}}^{\prime} \boldsymbol{x}_{\text{r}, f}^{\prime}
  -  \boldsymbol{s}_{\text{r}} +  f^{\prime}  \boldsymbol{H}_{\text{r, fixed}}
\boldsymbol{x}_{\text{r, fixed}}    }  }_2^2  
 +f^{\prime2} 
   \mathrm{E} \{  \boldsymbol{w}_{\text{r}}^T
   \boldsymbol{w}_{\text{r}}
  \}  \text{.}
\end{align}
 As in the previous subsection the Hessian can be constructed with the partial derivatives which are now given by  
 \begin{align}
 \label{eq:partial_2}
\boldsymbol{\Gamma}  & =
\frac{ \partial^2 J(\boldsymbol{x}_{\text{r}, f}^{\prime}, f^{\prime})}{ \partial  \boldsymbol{x}_{\text{r}, f}^{\prime}  \partial  \boldsymbol{x}_{\text{r}, f}^{\prime T}  } 
=
2 
   \boldsymbol{H}_{\text{r}}^{\prime T}
   \boldsymbol{H}_{\text{r}}^{\prime}  \text{,} \notag\\
\epsilon & =\frac{ \partial^2 J(\boldsymbol{x}_{\text{r}, f}^{\prime}, f^{\prime})}{ \partial  f^{\prime\ 2} }
=2 \lVert  
 \boldsymbol{H}_{\text{r, fixed}}
\boldsymbol{x}_{\text{r, fixed}}        \rVert_2^2  
 +
 2 
   \mathrm{E} \{  \boldsymbol{w}_{\text{r}}^T
   \boldsymbol{w}_{\text{r}}
  \}   \geq 0  \text{,}    \\ 
\boldsymbol{\eta} & =
\frac{ \partial^2 J(\boldsymbol{x}_{\text{r}, f}^{\prime}, f^{\prime})}{ \partial \boldsymbol{x}_{\text{r}, f}^{\prime}  \partial  f^{\prime} }
 =  
   2
   \boldsymbol{H}_{\text{r}}^{\prime T}
   \boldsymbol{H}_{\text{r, fixed}}
   \boldsymbol{x}_{\text{r, fixed}}  \text{.} \notag
   \end{align} 
Analogous to the previous subsection we assume that $\boldsymbol{\Gamma}\succ \boldsymbol{0}$ and then \eqref{eq:epsilon_ieq} is a sufficient condition for convexity. 
In this case, \eqref{eq:epsilon_ieq}, after including the partial derivatives \eqref{eq:partial_2},
can be rearranged to
\begin{align}
 &\mathrm{E} \{  \boldsymbol{w}_{\text{r}}^T
   \boldsymbol{w}_{\text{r}}
  \}   
  \geq \boldsymbol{x}_{\text{r, fixed}}^T
  \boldsymbol{H}_{\text{r, fixed}}^T
(    \boldsymbol{H}_{\text{r}}^{\prime}
     (  \boldsymbol{H}_{\text{r}}^{\prime T}
   \boldsymbol{H}_{\text{r}}^{\prime}  )^{-1}    
   \boldsymbol{H}_{\text{r}}^{\prime T} -\boldsymbol{I}     )
   \boldsymbol{H}_{\text{r, fixed}}
   \boldsymbol{x}_{\text{r, fixed}} \text{.}
\label{eq:convexity_condition}
\end{align}
Convexity is established by showing that the RHS of \eqref{eq:convexity_condition} is always smaller or equal to zero. This can be shown by considering
\begin{align}
  \boldsymbol{v}^T
  (    \boldsymbol{H}_{\text{r}}^{\prime}
     (  \boldsymbol{H}_{\text{r}}^{\prime T}
   \boldsymbol{H}_{\text{r}}^{\prime}  )^{-1}    
   \boldsymbol{H}_{\text{r}}^{\prime T} -\boldsymbol{I}     )
   \boldsymbol{v}    \leq 0  \text{,}
\end{align}
which holds for all $\boldsymbol{v}$, since
$ \boldsymbol{H}_{\text{r}}^{\prime}
     (  \boldsymbol{H}_{\text{r}}^{\prime T}
   \boldsymbol{H}_{\text{r}}^{\prime}  )^{-1}    
   \boldsymbol{H}_{\text{r}}^{\prime T}$ is a projection matrix where the eigenvalues can only be one or zero.

\subsection{Derivation of ${h_k^\text{eff}}$}
\label{subsec:derivation_of_heff}
In the following, the expression for $h_k^\text{eff}$ is derived. 
To minimize $\lambda_{\epsilon_k}^2$ the derivative of \eqref{eq:problem_for_heff} with respect to $\gamma^*$ is taken and equated to 0, which is expressed as
\begin{align}
     \frac{\partial{\lambda_{\epsilon_k}^2}}{\partial{\gamma^*}}&=\gamma \ \text{E}\chav{s_k(\boldsymbol{s})\ s_k(\boldsymbol{s})^*}-\text{E}\chav{\boldsymbol{h}_k\ \boldsymbol{x}(\boldsymbol{s})\ s_k(\boldsymbol{s})^*}=0.
\end{align}
With this, the effective channel coefficient reads as
\begin{align}
     h_k^\text{eff}&=\gamma= \frac{ \boldsymbol{h}_k\ \text{E}\chav{s_k(\boldsymbol{s})^*\ \boldsymbol{x}(\boldsymbol{s})}}{\sigma_s^2}.
 \end{align}
Finally, $h^\text{eff}_k$ is used for the computation of the mean squared error $\lambda_{\epsilon_k}^2=\text{E}\chav{ \PM{\epsilon_k[t]}^2}$ as follows
\begin{align}
         \lambda_{\epsilon_k}^2&= \boldsymbol{h}_k  \text{E}\chav{\boldsymbol{x}(\boldsymbol{s})\ \boldsymbol{x}(\boldsymbol{s})^H}\boldsymbol{h}_k^H - \ \frac{1}{\sigma_s^2}\PM{\text{E}\chav{\boldsymbol{h}_k\ \boldsymbol{x}(\boldsymbol{s})\ k(\boldsymbol{s})^*}}^2\notag\\
         &=\boldsymbol{h}_k\ \boldsymbol{\Lambda}_x \ \boldsymbol{h}^H_k-\PM{h^\text{eff}_k}^2 \sigma_s^2. 
\end{align}

\ifCLASSOPTIONcaptionsoff
  \newpage
\fi

\bibliographystyle{IEEEtran}
\bibliography{bib-refs}

\begin{thebibliography}{10}
\providecommand{\url}[1]{#1}
\csname url@samestyle\endcsname
\providecommand{\newblock}{\relax}
\providecommand{\bibinfo}[2]{#2}
\providecommand{\BIBentrySTDinterwordspacing}{\spaceskip=0pt\relax}
\providecommand{\BIBentryALTinterwordstretchfactor}{4}
\providecommand{\BIBentryALTinterwordspacing}{\spaceskip=\fontdimen2\font plus
\BIBentryALTinterwordstretchfactor\fontdimen3\font minus
  \fontdimen4\font\relax}
\providecommand{\BIBforeignlanguage}[2]{{%
\expandafter\ifx\csname l@#1\endcsname\relax
\typeout{** WARNING: IEEEtran.bst: No hyphenation pattern has been}%
\typeout{** loaded for the language `#1'. Using the pattern for}%
\typeout{** the default language instead.}%
\else
\language=\csname l@#1\endcsname
\fi
#2}}
\providecommand{\BIBdecl}{\relax}
\BIBdecl

\bibitem{MMSE_bb}
E.~S.~P. {Lopes} and L.~T.~N. {Landau}, ``{Optimal and Suboptimal MMSE
  Precoding for Multiuser MIMO Systems Using Constant Envelope Signals with
  Phase Quantization at the Transmitter and PSK Modulation},'' in \emph{WSA
  2020; 24th International ITG Workshop on Smart Antennas}, Hamburg, Germany,
  2020.

\bibitem{6G_Future_Directions}
L.~U. {Khan}, I.~{Yaqoob}, M.~{Imran}, Z.~{Han}, and C.~S. {Hong}, ``{6G
  Wireless Systems: A Vision, Architectural Elements, and Future Directions},''
  \emph{{{IEEE} Access}}, vol.~8, pp. 147\,029--147\,044, 2020.

\bibitem{Power_consumption}
F.~{Rusek}, D.~{Persson}, B.~K. {Lau}, E.~G. {Larsson}, T.~L. {Marzetta},
  O.~{Edfors}, and F.~{Tufvesson}, ``{Scaling Up MIMO: Opportunities and
  Challenges with Very Large Arrays},'' \emph{{{IEEE} Signal Process. Mag.}},
  vol.~30, no.~1, 2013.

\bibitem{6G_Use_cases}
M.~{Giordani}, M.~{Polese}, M.~{Mezzavilla}, S.~{Rangan}, and M.~{Zorzi},
  ``{Toward 6G Networks: Use Cases and Technologies},'' \emph{{{IEEE} Commun.
  Mag.}}, vol.~58, no.~3, pp. 55--61, 2020.

\bibitem{6G_research_challanges}
S.~{Elmeadawy} and R.~M. {Shubair}, ``{6G Wireless Communications: Future
  Technologies and Research Challenges},'' in \emph{2019 Int. Conference on
  Electrical and Computing Technologies and Applications (ICECTA)}, Ras Al
  Khaimah, UAE, 2019.

\bibitem{6G_Vision}
P.~{Yang}, Y.~{Xiao}, M.~{Xiao}, and S.~{Li}, ``{6G Wireless Communications:
  Vision and Potential Techniques},'' \emph{{{IEEE} Netw.}}, vol.~33, no.~4,
  pp. 70--75, 2019.

\bibitem{Amine_2011power}
A.~Mezghani and J.~A. Nossek, ``Power efficiency in communication systems from
  a circuit perspective,'' in \emph{{2011 IEEE International Symposium of
  Circuits and Systems (ISCAS)}}, 2011, pp. 1896--1899.

\bibitem{Walden_1999}
R.~Walden, ``Analog-to-digital converter survey and analysis,'' \emph{IEEE J.
  Sel. Areas Commun.}, vol.~17, no.~4, Apr. 1999.

\bibitem{ZF-Precoding}
S.~K. {Mohammed} and E.~G. {Larsson}, ``{Per-Antenna Constant Envelope
  Precoding for Large Multi-User MIMO Systems},'' \emph{{{IEEE} Trans.
  Commun.}}, vol.~61, no.~3, pp. 1059--1071, March 2013.

\bibitem{Jacobsson_2017}
S.~{Jacobsson}, G.~{Durisi}, M.~{Coldrey}, T.~{Goldstein}, and C.~{Studer},
  ``{Quantized Precoding for Massive MU-MIMO},'' \emph{{{IEEE} Trans.
  Commun.}}, vol.~65, no.~11, pp. 4670--4684, 2017.

\bibitem{M_Joham_ZF}
M.~{Joham}, W.~{Utschick}, and J.~A. {Nossek}, ``Linear transmit processing in
  {MIMO} communications systems,'' \emph{{{IEEE} Trans. Signal Process.}},
  vol.~53, no.~8, pp. 2700--2712, Aug 2005.

\bibitem{Mezghani2009}
A.~{Mezghani}, R.~{Ghiat}, and J.~A. {Nossek}, ``Transmit processing with low
  resolution {D/A}-converters,'' in \emph{2009 16th {IEEE} Int. Conference on
  Electronics, Circuits and Systems - ({ICECS} 2009)}, Hammamet, Tunisia, Dec
  2009, pp. 683--686.

\bibitem{one_bit_zf_saxena}
A.~K. {Saxena}, I.~{Fijalkow}, and A.~L. {Swindlehurst}, ``On one-bit quantized
  {ZF} precoding for the multiuser massive {MIMO} downlink,'' in \emph{2016
  {IEEE} Sensor Array and Multichannel Signal Processing Workshop ({SAM})}, Rio
  de Janeiro, Brazil, July 2016.

\bibitem{Squid_precoder}
S.~{Jacobsson}, G.~{Durisi}, M.~{Coldrey}, T.~{Goldstein}, and C.~{Studer},
  ``{Nonlinear 1-bit precoding for massive MU-MIMO with higher-order
  modulation},'' in \emph{2016 50th Asilomar Conference on Signals, Systems and
  Computers}, 2016, pp. 763--767.

\bibitem{CVX-CIO}
P.~V. {Amadori} and C.~{Masouros}, ``Constant envelope precoding by
  interference exploitation in phase shift keying-modulated multiuser
  transmission,'' \emph{{{IEEE} Trans. Commun.}}, vol.~16, no.~1, pp. 538--550,
  Jan 2017.

\bibitem{Jedda_mmse_mapped}
A.~{Noll}, H.~{Jedda}, and J.~{Nossek}, ``{PSK Precoding in Multi-User MISO
  Systems},'' in \emph{WSA 2017; 21th International ITG Workshop on Smart
  Antennas}, Berlin, Germany, 2017, pp. 1--7.

\bibitem{MSM_precoder}
H.~{Jedda}, A.~{Mezghani}, A.~L. {Swindlehurst}, and J.~A. {Nossek},
  ``Quantized constant envelope precoding with {PSK} and {QAM} signaling,''
  \emph{{{{IEEE} Trans. Wireless Commun.}}}, vol.~17, no.~12, pp. 8022--8034,
  Dec 2018.

\bibitem{L_Chu2019}
L.~{Chu}, F.~{Wen}, L.~{Li}, and R.~{Qiu}, ``{Efficient Nonlinear Precoding for
  Massive MIMO Downlink Systems With 1-Bit DACs},'' \emph{{{{IEEE} Trans.
  Wireless Commun.}}}, vol.~18, no.~9, pp. 4213--4224, 2019.

\bibitem{jacobsson2018nonlinear}
S.~Jacobsson, O.~Casta{\~n}eda, C.~Jeon, G.~Durisi, and C.~Studer, ``{Nonlinear
  precoding for phase-quantized constant-envelope massive MU-MIMO-OFDM},'' in
  \emph{2018 25th International Conference on Telecommunications (ICT)}, 2018.

\bibitem{Magiq}
A.~{Nedelcu}, F.~{Steiner}, M.~{Staudacher}, G.~{Kramer}, W.~{Zirwas}, R.~S.
  {Ganesan}, P.~{Baracca}, and S.~{Wesemann}, ``Quantized precoding for
  multi-antenna downlink channels with {MAGIQ},'' in \emph{Proc. of the 22nd
  Int. ITG Workshop on Smart Antennas}, Bochum, Germany, March 2018.

\bibitem{Struder_c3po}
O.~Castañeda, S.~Jacobsson, G.~Durisi, T.~Goldstein, and C.~Studer, ``{VLSI
  Design of a 3-bit Constant-Modulus Precoder for Massive MU-MIMO},'' in
  \emph{2018 IEEE International Symposium on Circuits and Systems (ISCAS)},
  2018, pp. 1--5.

\bibitem{Landau2017}
L.~T.~N. {Landau} and R.~C. {de Lamare}, ``Branch-and-bound precoding for
  multiuser {MIMO} systems with 1-bit quantization,'' \emph{{{IEEE} Wireless
  Commun. Lett.}}, vol.~6, no.~6, pp. 770--773, Dec 2017.

\bibitem{General_MMDDT_BB}
E.~S.~P. {Lopes} and L.~T.~N. {Landau}, ``{Optimal Precoding for Multiuser MIMO
  Systems With Phase Quantization and PSK Modulation via Branch-and-Bound},''
  \emph{{{IEEE} Wireless Commun. Lett.}}, vol.~9, no.~9, pp. 1393--1397, 2020.

\bibitem{Jacobsson2018}
S.~{Jacobsson}, W.~{Xu}, G.~{Durisi}, and C.~{Studer}, ``{MSE}-optimal 1-bit
  precoding for multiuser {MIMO} via branch and bound,'' in \emph{Proc. {IEEE}
  Int. Conf. Acoust., Speech, Signal Process.}, Calgary, Alberta, Canada, April
  2018, pp. 3589--3593.

\bibitem{ML_one_bit}
A.~{Mezghani}, M.~{Khoufi}, and J.~A. {Nossek}, ``{Maximum likelihood detection
  for quantized MIMO systems},'' in \emph{2008 International ITG Workshop on
  Smart Antennas}, Vienna, Austria, 2008, pp. 278--284.

\bibitem{mimo_1bit_detection}
\BIBentryALTinterwordspacing
C.~Risi, D.~Persson, and E.~G. Larsson, ``Massive {MIMO} with 1-bit {ADC},''
  \emph{CoRR}, vol. abs/1404.7736, 2014. [Online]. Available:
  \url{http://arxiv.org/abs/1404.7736}
\BIBentrySTDinterwordspacing

\bibitem{One_bit_sphere_decoding}
Y.~{Jeon}, N.~{Lee}, S.~{Hong}, and R.~W. {Heath}, ``{One-Bit Sphere Decoding
  for Uplink Massive MIMO Systems With One-Bit ADCs},'' \emph{{{{IEEE} Trans.
  Wireless Commun.}}}, vol.~17, no.~7, pp. 4509--4521, 2018.

\bibitem{Z_shao_2018}
Z.~{Shao}, R.~C. {de Lamare}, and L.~T.~N. {Landau}, ``{Iterative Detection and
  Decoding for Large-Scale Multiple-Antenna Systems With 1-Bit ADCs},''
  \emph{{{IEEE} Wireless Commun. Lett.}}, vol.~7, no.~3, pp. 476--479, 2018.

\bibitem{Z_Zhang2018}
Z.~{Zhang}, X.~{Cai}, C.~{Li}, C.~{Zhong}, and H.~{Dai}, ``{One-Bit Quantized
  Massive MIMO Detection Based on Variational Approximate Message Passing},''
  \emph{{{IEEE} Trans. Signal Process.}}, vol.~66, no.~9, pp. 2358--2373, 2018.

\bibitem{J_choi2016}
J.~{Choi}, J.~{Mo}, and R.~W. {Heath}, ``{Near Maximum-Likelihood Detector and
  Channel Estimator for Uplink Multiuser Massive MIMO Systems With One-Bit
  ADCs},'' \emph{{{IEEE} Trans. Commun.}}, vol.~64, no.~5, pp. 2005--2018,
  2016.

\bibitem{Wang2015}
S.~{Wang}, Y.~{Li}, and J.~{Wang}, ``{Multiuser Detection in Massive Spatial
  Modulation MIMO With Low-Resolution ADCs},'' \emph{{{{IEEE} Trans. Wireless
  Commun.}}}, vol.~14, no.~4, pp. 2156--2168, 2015.

\bibitem{Studer_NN}
A.~Balatsoukas-Stimming, O.~Castañeda, S.~Jacobsson, G.~Durisi, and C.~Studer,
  ``Neural-network optimized 1-bit precoding for massive {MU-MIMO},'' in
  \emph{2019 IEEE 20th International Workshop on Signal Processing Advances in
  Wireless Communications (SPAWC)}, Cannes, France, 2019.

\bibitem{Boyd_2004}
S.~Boyd and L.~Vandenberghe, \emph{Convex Optimization}.\hskip 1em plus 0.5em
  minus 0.4em\relax New York, NY, USA: Cambridge University Press, 2004.

\bibitem{Jedda_2016}
H.~Jedda, J.~A. Nossek, and A.~Mezghani, ``Minimum {BER} precoding in 1-bit
  massive {MIMO} systems,'' in \emph{Proc. of IEEE Sensor Array and
  Multichannel Signal Processing Workshop (SAM)}, Rio de Janeiro, Brazil, July
  2016.

\bibitem{Ten_brink_idd}
S.~{ten Brink}, J.~{Speidel}, and {Ran-Hong Yan}, ``Iterative demapping and
  decoding for multilevel modulation,'' in \emph{IEEE GLOBECOM 1998}, vol.~1,
  Sydney, Australia, 1998, pp. 579--584 vol.1.

\bibitem{Marzetta_2013}
H.~{Yang} and T.~L. {Marzetta}, ``Performance of conjugate and zero-forcing
  beamforming in large-scale antenna systems,'' \emph{{{IEEE} J. Sel. Areas
  Commun.}}, vol.~31, no.~2, pp. 172--179, 2013.

\bibitem{SPA-Decoder}
F.~R. {Kschischang}, B.~J. {Frey}, and H.~A. {Loeliger}, ``Factor graphs and
  the sum-product algorithm,'' \emph{{{IEEE} Trans. Inf. Theory}}, vol.~47,
  no.~2, pp. 498--519, 2001.

\end{thebibliography}
\end{document}